\documentclass[useAMS,usenatbib,letterpaper]{mn2e}
\usepackage{graphicx, amsmath, amssymb}
\usepackage{journal_abbreviations, custom_commands}
\usepackage[totalwidth=500pt,totalheight=650pt,centering,layoutvoffset=-0.3cm]{geometry}

\def\ngaltotal{854}

\def\ngaltotalSS{21}
\def\ngaluvSS{4}
\def\ngalnebSS{17}

\def\wa{0.5\textwidth}

\def\wc{0.33\textwidth}

\title[Detection of hot, metal-enriched outflowing gas]
{Detection of hot, metal-enriched outflowing gas
around \textit{z}$\, \approx\,$2.3 star-forming 
galaxies in the Keck Baryonic Structure Survey}
\author[Turner et al.]{Monica L. Turner,$^{1}$\thanks{E-mail: turnerm@strw.leidenuniv.nl}
Joop Schaye$^{1}$,
Charles C. Steidel$^{2}$,\newauthor
Gwen C. Rudie$^{3}$,
and Allison L. Strom$^{2}$\\
$^{1}$Leiden Observatory, Leiden University, PO Box 9513, 2300 RA Leiden, The Netherlands\\
$^{2}$California Institute of Technology, MS 249-17, Pasadena, CA 91125, USA\\
$^{3}$Carnegie Observatories, 813 Santa Barbara Street, Pasadena, CA 91101, USA\\
}

\begin{document}	

\date{\today}

\pagerange{\pageref{firstpage}--\pageref{lastpage}}

\maketitle

\label{firstpage}

\begin{abstract}
We use quasar absorption lines to study the physical conditions in the circumgalactic 
medium of redshift $z\approx 2.3$ star-forming galaxies taken from the Keck Baryonic Structure Survey (KBSS).  
In \citet{turner14} we used the pixel optical depth technique to show that absorption 
by \hone\ and the metal ions \osix, \nfive, \cfour, \cthree\ and \sifour\ is 
strongly enhanced within $|\Delta v|\lesssim170$~\kmps\ and 
projected distances $|d|\lesssim180$ proper kpc
from sightlines to the background quasars. 
Here we demonstrate that the \osix\ absorption is also strongly enhanced at fixed \hone, \cfour, 
and \sifour\ optical depths, and that this enhancement extends out to 
$\sim350$~\kmps. At fixed \hone\ the increase in the median \osix\ optical depth near 
galaxies is 0.3--0.7 dex and is detected at 2--3-$\sigma$ confidence for all seven \hone\ bins
that have $\log_{10}\tau_{\honem}\ge-1.5$. We use ionization 
models to show that the observed strength of \osix\ as a function of \hone\ 
is consistent with enriched, photoionized gas for pixels with $\tau_{\honem} \gtrsim 10$.
However, for pixels with $\tau_{\honem} \lesssim 1$ this would lead to 
implausibly high metallicities at low densities if the gas were photoionized by the background 
radiation. This indicates that the galaxies are surrounded by gas that is sufficiently hot to be collisionally ionized
($T > 10^5\,$K) and that a substantial fraction of the hot gas has a metallicity $\gtrsim 10^{-1}$ of solar. 
Given the high metallicity and large velocity extent (out to $\sim1.5\times v_{\text{circ}}$) of this gas, 
we conclude that we have detected hot, metal enriched outflows arising from star-forming galaxies. 	
\end{abstract}



\section{Introduction}
\label{sec:intro}

Galaxy formation models predict that massive galaxies are surrounded by haloes of hot, 
chemically enriched gas, which may be penetrated by accreting streams of cold and largely metal-poor gas 
\citep[e.g.][]{keres05,dekel09,vandevoort11a,vandevoort12}. The hot component is heated through shocks 
associated with galactic winds and gas accretion, while the cold component is photo-heated  to 
temperatures of $T\sim 10^4\,$K. The circumgalactic medium thus results from the inflow of gas into the 
potential well set by the dark matter halo, and the outflows driven by feedback from star formation and/or 
active galactic nuclei (AGN). Hence, the gas around galaxies holds valuable clues 
to the fuelling and feedback processes that 
currently limit our understanding of galaxy evolution.

This  circumgalactic gas is very diffuse, making it difficult to detect in emission.
Instead, the gas can be studied in absorption using high-quality spectra of background quasars. However, the low
number density of bright quasars and the difficulty of obtaining accurate galaxy redshifts make it 
challenging to apply this technique to systematic surveys of the circumgalactic medium. 

The advent of the Cosmic Origin Spectrograph on the Hubble Space Telescope and the MOSFIRE spectrograph 
on the Keck telescope have recently improved the situation by significantly increasing the quality of 
low-redshift quasar spectra and the accuracy with which redshifts can be measured for high-redshift 
galaxies, respectively. In particular, \citet{tumlinson11} found \osix\ to be ubiquitous  within 150~proper 
kiloparsec (pkpc) of $z=0.10$--$0.36$ star-forming galaxies  with median halo masses 
$\approx1.6\times10^{12}$~${\rm M}_{\odot}$ \citep{werk14}. At $z\approx 2.4$, 
\citet{turner14} found strong enhancements of \osix, \nfive, \cfour, \cthree, \sifour, and \hone\ within 
180~pkpc and 240 km$\,$s$^{-1}$ of star-forming galaxies that are also thought to be hosted by haloes with 
masses $\sim 10^{12}\,{\rm M}_\odot$. These studies confirmed and extended 
earlier work on the association between galaxies and absorbers.

The ion \osix\ is of particular interest, because its absorption is relatively strong and because it can 
trace gas at $T\sim10^5$--$10^6$~K, similar to the temperatures to which the gas is 
expected to be heated in shocks associated with winds and accretion events. Furthermore, 
it is the temperature range for which gas cooling from much higher temperatures is most
likely to be detectable. 
Indeed, a number of simulations predict the presence of large amounts of
\osix\ around $M_{\rm halo}>10^{11} {\rm M}_{\odot}$ 
galaxies \citep[e.g.][]{stinson12, ford13, shen13}.
In general, these studies find that \osix\ resides in a collisionally ionized gas
phase for impact parameters  $<100$~pkpc, outside of which \osix\ tends to be found in a cooler, photoionized gas.
\citet{tepper-garcia11} estimate that two-thirds of \osix\ absorbers 
at $z=0.25$ in their simulations trace gas with $T>10^5$~K (see also \citealt{oppenheimer12}). 

At low redshifts, many observational programmes have found that \osix\ absorbers 
tend to be located within 300~pkpc of galaxy positions 
\citep[e.g.][]{stocke06, chen09, prochaska11, tumlinson11}.
Although the presence of \osix\ around galaxies is well established, 
the properties of the gas that it traces are still under debate, 
as \osix\ systems are observed to arise in both photoionized 
and collisionally ionized phases \citep[e.g,][]{danforth08, savage14}. 
While some studies find that \osix\ in galaxy haloes is consistent with being photoionized
\citep[e.g.][]{prochaska11},
many circumgalactic \osix\ absorbers are thought to be  in a phase distinct
from that of both \hone\ and lower ions at $T<10^5$~K, due to their
often complex and differing kinematic structures \citep[e.g.][]{tripp08}
and/or inferences from ionization modelling \citep[e.g.][]{werk13}.
Using \hone\ \lya\ absorber stacking\footnote{Although this work does 
not use direct galaxy detections, the authors inferred that
their strongest absorber sample probes regions defined as circumgalactic
in \citet{rudie12} in $\approx60$ percent of cases. This was determined
by using high-resolution QSO spectra from fields where the redshifts of Lyman break galaxies
with impact parameters $<300$~kpc are already known. The QSO spectra were 
degraded to match the resolution of the spectra used in \citet{pieri14}, and a correction
for the expected volume density of Lyman break galaxies from \citet{reddy08} was applied.} 
\citet{pieri14} found that for high ionization lines including \osix,
their observations were most consistent with $T=10^4\text{--}10^{4.5}$~K,
with a possible contribution from warmer
$T\sim10^{5.5}$~K gas (inferred from N/O measurements)	. 

Because of its near-coincidence with the \hone\ \lyb\ forest, 
studying \osix\ absorption using line-fitting techniques
becomes progressively more difficult with increasing redshift,
although individual line fitting is still possible and has been 
done extensively at $z>2$ \citep[e.g.][]{bergeron02, carswell02, 
simcoe02, simcoe04, simcoe06, lopez07, schaye07, fox08}. Rather than looking by eye 
for individual \osix\ absorption lines, \osix\ has also been studied using the automatic 
pixel optical depth technique to correlate the absorption from \osix\ with that of other 
ions \citep{cowie98, schaye00a, aguirre08, pieri10}, optionally correcting for 
much of the contamination by \hone\ Lyman series lines \citep{aguirre02,aguirre08,turner14}. 

\citet{aracil04} divided a sample of pixel optical depths into those near and far 
from strong \hone\ \lya\ absorption. 
They found that for gas at the same \hone\ optical depth, the amount
of \osix\ was enhanced near strong \hone\ \lya\ absorption for $2.1<z<3.2$. 
A similar study was then undertaken by \citet{pieri06}
for the quasar Q$1422+231$ ($z=3.62$). Motivated by the very strong correlation 
between galaxies and strong \cfour\ absorbers found by \citet{adelberger03, adelberger05b}, 
they used \cfour\ optical depth as a galaxy proxy,
and found enhancements in both \cfour\ and \osix\ at fixed 
\hone\  close to galaxy positions. 

In this work we extend the technique used in \citet{pieri06} and apply 
it to a much larger sample of 15 quasars and \ngaltotal\ spectroscopically 
confirmed galaxies (with impact parameters 
as small as 40~pkpc) taken from the Keck Baryonic Structure Survey 
(KBSS, \citealt{steidel14}). \citet{rakic12} and \citet{rudie12} 
have already used an earlier version of the survey to study the 
distribution of neutral hydrogen around the galaxies, 
while \citet{turner14} measured the distribution of metal ions 
using the same data as is analysed here. Using a galactocentric 
approach and the pixel optical depth technique, \citet{turner14} 
found metal-line absorption to be strongly enhanced with respect to random regions for impact parameters
$\lesssim180$~pkpc and line of sight (LOS) distances within $\pm240$~\kmps\ of the galaxy 
positions (or $\sim$1~pMpc in the case of pure Hubble flow). 
Furthermore, thanks to observations using MOSFIRE,
the elongation of enhancement along the LOS was determined to be
largely caused by gas peculiar velocities (rather than redshift errors).

While \citet{turner14} studied optical depth as a function of galaxy 
distance, here we will measure the enhancement of metal-line absorption at fixed \hone\ optical depth. 
This enables us to tell whether the enhancement in metal absorption near galaxies found by \citet{turner14} 
merely reflects the higher gas densities implied by the observed increase in \hone\ absorption, or whether 
it indicates that the circumgalactic gas has a higher metallicity or a different 
temperature compared to random regions with the same \hone\ optical depth. 

This paper is organized as follows. In Section~\ref{sec:method} we briefly review the properties of the 
galaxy and quasar samples, the galaxy redshifts, and the pixel optical depth technique. In Section~\ref{sec:results} 
we present the principal observational results of this paper, which is that 
we find a strong and significant enhancement of \osix\ at fixed 
\hone, \cfour\ and \sifour\ for impact parameters $<180$~pkpc and velocities $\lesssim350$~\kmps. 
In section~\ref{sec:models} we consider whether the 
observational result can be explained 
(1) if the gas near galaxies is photoionized and metal rich, 
(2) if the gas is photoionized by radiation from stars in the nearby galaxies,
or (3) if the enriched gas is collisionally ionized.
We find that small galactocentric distance pixels  with $\tau_{\honem} \gtrsim 10$ are 
in agreement with scenarios (1) and (3), while for those 
that have $\tau_{\honem} \lesssim 1$,
only scenario (3), i.e.\ the presence of hot, collisionally ionized gas, 
provides a consistent explanation.
Finally, in 
Section~\ref{sec:conclusion} we summarize and discuss our main results.

Throughout the paper, we use proper rather than comoving units
(denoted as pkpc and pMpc), and employ cosmological parameters
as measured from the Planck mission \citep{planck13}, 
i.e.\ $H_{\rm 0}=67.1$~\kmps~Mpc$^{-1}$, $\Omega_{\rm m} = 0.318$, $\Omega_{\Lambda} = 0.683$, 
and $\Omega_{\rm b}h^2=0.0221$.


\section{Observations and method}
\label{sec:method}

This work makes use of procedures that were detailed in
\citet{turner14}, which we will briefly outline here.

\begin{table}
\caption{The log of the median optical depth, and the median 
  continuum S/N of all pixels (with normalized flux $>0.7$)
   in the redshift range considered for 
  the particular ion and recovery method.}
\label{tab:taumed}
\begin{tabular}{@{}lcc}
\hline\hline
Ion & $\log_{10}$\taumedk  & S/N$_{\rm med}$\\
\hline
\hone & $-1.29$ & $70.1$ \\ 
\osix & $-1.64$ & $45.7$ \\ 
\cfour & $-2.95$ & $83.6$ \\ 
\sifour & $-3.16$ & $84.4$ \\ 
\hline
\end{tabular}

\end{table}

\subsection{Galaxy \& QSO samples}
\label{sec:sample}

The KBSS is centred around the fields of 15 hyper-luminous
QSOs, all of which have been observed extensively with 
Keck/HIRES and therefore have very high quality spectra. 
Details about the QSO reduction and analysis, including the fitting
out of DLA wings, are described in \citet{rudie12}. 
The HIRES spectra typically have $R\simeq45000$ (which corresponds 
to a $\text{FWHM}\simeq7$~\kmps), and S/N ranging from $\sim50$ to $200$~pixel$^{-1}$
(we give the median S/N for the spectral regions covered 
by each ion in Table~\ref{tab:taumed}).

The survey focuses on obtaining spectroscopic redshifts 
for the galaxies in the above QSO fields. The full sample currently consists of  
$\approx2550$ galaxies at $\langle z\rangle\approx2.3$,
that were  chosen to have redshifts in the range probed by the QSO
spectra using UV-colour selection techniques from 
\citet{steidel03, steidel04} and \citet{adelberger04}. 
Spectroscopic follow-up using the instruments LRIS, 
NIRSPEC and/or MOSFIRE was then performed on galaxies with 
apparent magnitudes $m_\mathcal{R}\leq25.5$ (see \citealt{rudie12}
for more information about the galaxy follow-up strategy). 
The above selection typically results in galaxies with
halo masses $\sim10^{12}$~\msol
\citep{adelberger05a, conroy08, trainor12, rakic13},
which corresponds to virial radii and circular velocities
of $\approx90$~pkpc and $\approx217$~\kmps, respectively.
Furthermore, these galaxies tend to have dynamical masses
$\approx7\times10^{10}$~\msol \citep{erb06c},
median star formation rates $\approx25$~\msol~yr$^{-1}$
\citep{erb06b, steidel14},
gas-phase metallicities $\approx0.4\,\text{Z}_\odot$ \citep{steidel14},
and  stellar ages $\approx0.7$~Gyr \citep{erb06c}. 
 
In this work, we focus on the subset of 21 KBSS galaxies that satisfy the following two
constraints. First, we limit the sample to galaxies with impact parameters $<180$~pkpc. Although 
we also considered galaxies with impact parameters up to $2$~pMpc as was done in \citet{turner14},
we did not find any differences in the results from using impact parameter 
bins $>180$~pkpc compared to those from random regions. 
The choice of 180~pkpc can be further motivated by the fact that \citet{turner14}
found a strong metal optical depth enhancement above the median value for random regions up to the same 
impact parameter values. 

The second constraint concerns the velocity direction, where
we consider only galaxies that lie within the \lya\ 
forest of the background QSO, defined as:
\begin{equation}
  (1 + \zqsom) \dfrac{\lambda_{\lybm} }{ \lambda_{\lyam}}- 1  \leq z \leq \zqsom - (1+\zqsom) \dfrac{3000\,\kmpsm}{c}.
 \label{eq:zlim}
\end{equation}
The left-hand side corresponds to the beginning of the \lyb\ forest, while
the right-hand side is set by the QSO emission redshift 
less an offset factor to avoid proximity effects. 
This second criterion is required so that the pixel
optical depth analysis (as described in \S~\ref{sec:pod}) can be
applied to the same redshift range for different ions. 
 
\subsection{Galaxy redshifts}
\label{sec:galz}

NIRSPEC and MOSFIRE are both near-IR spectrographs, 
with galaxy redshift measurement uncertainties estimated to be $\Delta v \approx 60$ 
and $18$~\kmps, respectively. With these instruments, we are able to probe
the rest-frame optical wavelengths for the galaxies in this sample, 
and to observe the nebular emission lines H$\alpha$, H$\beta$, and
[\othree] $\lambda\lambda$4959,5007. Since these lines arise 
in \htwo\ regions of galaxies and, unlike Ly$\alpha$, are not subject to significant resonant scattering, they are thought to be robust tracers
of the systemic redshift, and for our analysis we therefore take the 
galaxy redshift to be equal to that of the nebular emission lines,
$z_{\rm gal} = z_{\rm neb}$. 

Although we are continuously working to increase the number of galaxies 
that have MOSFIRE observations, currently \ngaluvSS\ out of the \ngaltotalSS\ galaxies 
in our sub-sample have been observed only with LRIS, which probes the rest-frame UV. 
From these data, we can measure 
the galaxy redshifts from either interstellar absorption lines ($z_{\rm ISM}$) 
or \lya\ emission lines ($z_{\lyam}$). However,
these lines tend to be systematically offset from the systemic galaxy 
redshifts \citep{shapley03, adelberger03, steidel10, rakic11}. To correct for this,
we consider the galaxies that have both a nebular and rest-frame UV redshift 
measurement, and use the average difference between 
$z_{\rm ISM}$ or $z_{\lyam}$ and $z_{\rm neb}$ to determine a bulk correction value. 
The specifics of the correction implementation, as well as the latest
offset values, can be found in \S~2.2 of \citet{turner14}.
Our final sample of \ngaltotalSS\ galaxies 
contains \ngalnebSS\ and \ngaluvSS\ galaxies with redshifts
measured from nebular and rest-frame UV features, respectively. Their 
minimum, median, and maximum impact parameters are 35, 118,  
and 177~pkpc, respectively.

\subsection{Pixel optical depths}
\label{sec:pod}

In this work, we use the pixel optical depth method
\citep{cowie98, ellison00, schaye00a, aguirre02, schaye03, turner14}
 to study how absorption varies with galaxy proximity. 
Because of the statistical nature of our approach, we lose information about individual systems.
However, by studying correlations between the pixel optical depths of different transitions, we are able
to probe gas to lower densities, even in the presence of strong contamination, in a fast and fully automated manner.

A complete description of the method used is 
given in \S~3 and Appendix~A of \citet{turner14}; here
we give a short summary. 
We initially define the optical depth for ion $Z$ and multiplet component $k$ as 
\begin{equation}
 \tau_{Z,k}(z) \equiv -\ln[F_{Z,k}(z)]
\end{equation}
where $F_{Z,k}(z)$ is the normalized flux at the pixel with a wavelength given by
$\lambda = \lambda_{Z,k} [1 + z]$ where $\lambda_{Z,k}$ is the rest-wavelength 
of the $k$th transition of ion $Z$.

Next, we correct the absorption by each ion for saturation and for possible sources of contamination.
Beginning with \hone, if \lya\ is saturated, we use the optical depths
of higher-order transitions  (\lyb, \lyg\ etc.) at the same redshift. Of all the higher-order pixels 
at a single redshift, we take the minimum of the optical depths that are not saturated
(if there are any), scaled to the \lya\ transition, and use it to replace the saturated \lya\ value. 

The recovered \hone\ \lya\ is then used to subtract 5 orders of the Lyman series of \hone\ (starting from \lyb) 
from the optical  depths of ions that have rest-wavelengths in the \lyb\ forest: \osix\ $\lambda\lambda1032,1038$ 
and \cthree\ $977$. 
We note that for the recovery of ions that require \hone\ subtraction, 
we do not mask the DLAs or fit out the wings for the \hone\ recovery
(see \citealt{turner14}).

For ions that have a closely-spaced doublet (\osix\ $\lambda\lambda1032,1038$
and \sifour\ $\lambda\lambda1394,1403$)
we use the relative oscillator strengths to scale the optical depth of the weaker 
component to that of the stronger one, and take the minimum of the two optical depths
 at every redshift in order to correct for contamination. 

 Although \cfour\ $\lambda\lambda1548,1551$ also has a closely-spaced doublet, 
 due to its strength and the fact that it is located redwards of the \lya\ forest, 
 most contamination comes from other \cfour\ systems. We therefore use a procedure where we iteratively
 subtract the optical depth of the weaker component at the position of the stronger component. 
 
Although we attempt to correct for contamination, these corrections will generally not be 
perfect. In particular, we cannot correct for contamination of \osix\ by \hone\ \lya, which 
will cause our measurements of the \osix\ optical depth to be overestimates (however, on average
it will affect all pixels in the forest equally). This contaminating 
\lya\ absorption, as well as residual absorption from other contaminating lines, set the limit 
down to which we can detect enhancements in the \osix\ optical depth.


\section{Results}
\label{sec:results}

The first step in our analysis is to compute the median $\tau_{\osixm}$ as 
a function of $\tau_{\honem}$, which we will often denote \osix(\hone) for brevity. 
To do this, we take pixel pairs of $\tau_{\osixm}$ and $\tau_{\honem}$
at each redshift $z$, divide the pixels into 0.5~dex sized bins of 
$\log_{10}\tau_{\honem}$, and compute the median of 
$\tau_{\osixm}$ in each bin. 
We do this for both the full pixel sample (i.e.\ every pixel pair available
from all 15 QSOs in the redshift range given by eq.~\ref{eq:zlim}, 
irrespective of the locations of the galaxies), and for pixels known to be 
located at small galactocentric 
distances, defined as those within\footnote{This
velocity interval was chosen because (as explained in more detail in
\citealt{rakic12} and \citealt{turner14}) it is the scale over 
which the optical depths are smooth in the LOS direction.} $\pm170$~\kmps of 
the redshifts of galaxies with impact parameters $<180$~pkpc.

To avoid effects due to small number statistics, we do the following. 
For the full pixel sample, we divide the spectra into chunks of 5\AA.
We then compute the number of chunks that are sampled
by each $\log_{10}\tau_{\honem}$ bin, and discard any bins that draw from fewer
than five different chunks. 
For the small galactocentric distance pixel sample, we 
remove bins that do not have pixel contributions from at least five different galaxies.
Finally, for both pixel samples we discard any bins containing fewer 
than 25 pixels in total.

\begin{figure*}
 \includegraphics[width=\wc]{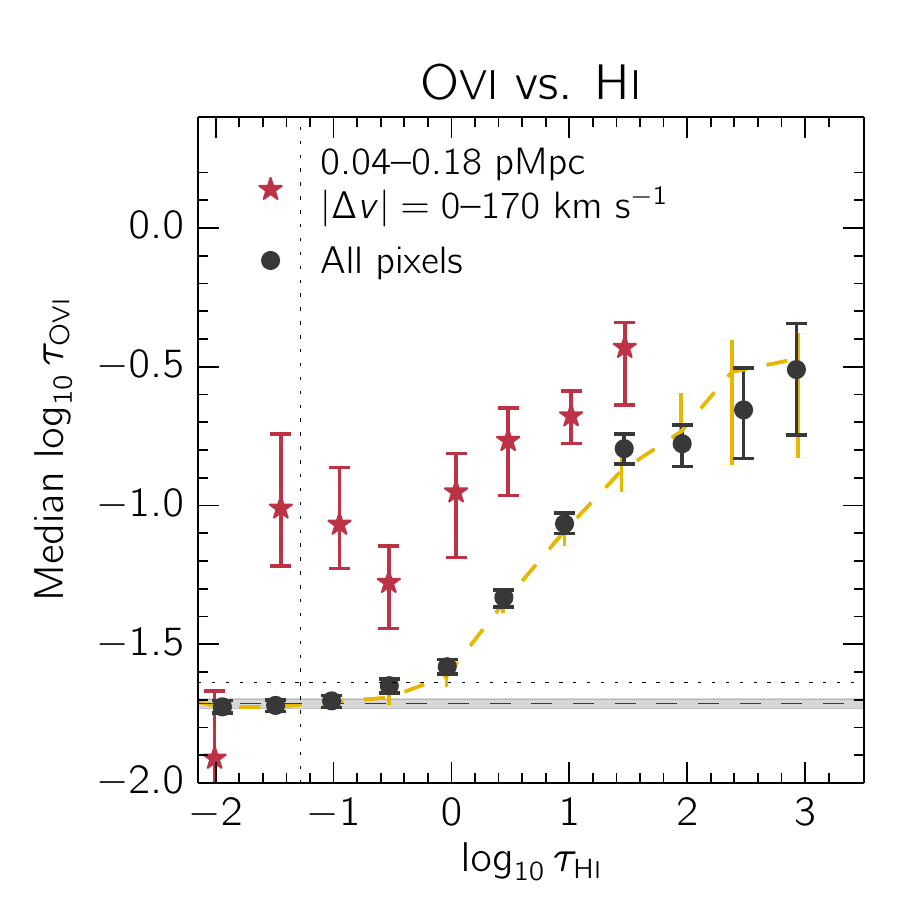} 
  \includegraphics[width=\wc]{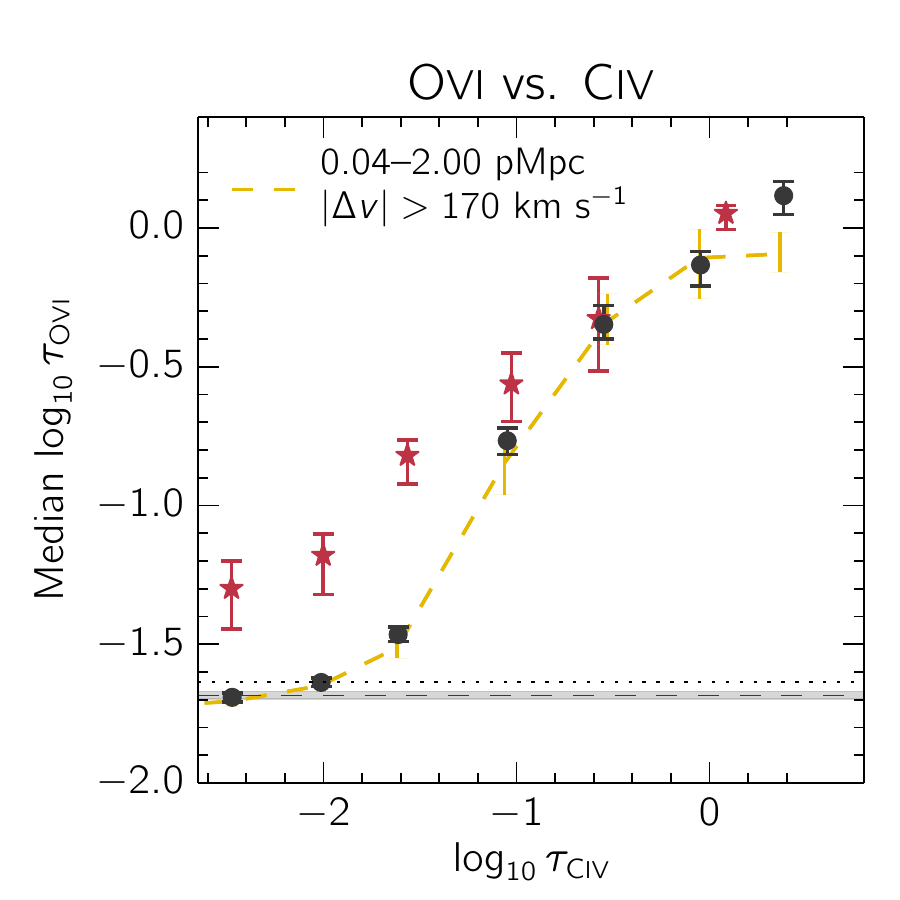} 
 \includegraphics[width=\wc]{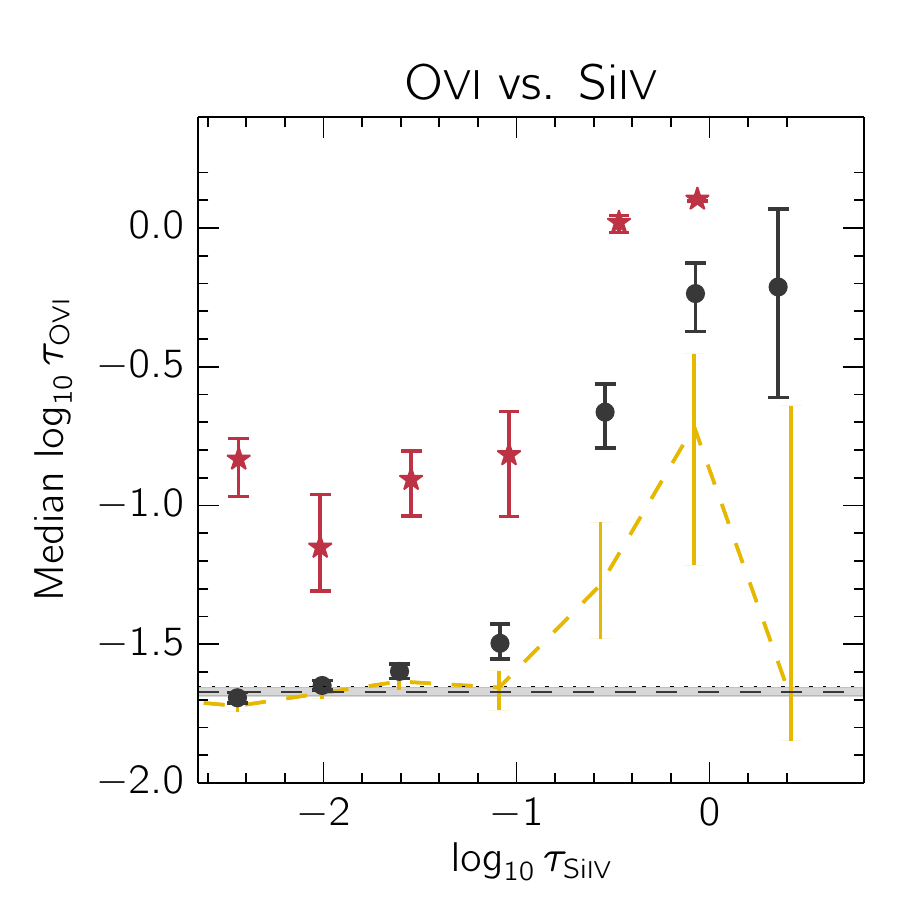} \\
 \caption{\textit{Left panel:} The \osix(\hone) relation for the full sample of 
  pixels (black circles) and for 
  pixels located at small galactocentric distances (red stars). 
  Error bars indicate the $1\sigma$ uncertainty determined by 
  bootstrap resampling the galaxies. The median values
  for the full sample are denoted by the horizontal and vertical
  dotted black lines. The \osix\ flat level,
  defined as the  median value for all \osix\ pixels that have associated 
 $\tau_{\honem}<\tau_{\honem,{\rm cut}}=0.1$,  is represented by the horizontal dashed
 grey line, and the shaded region shows its 1-$\sigma$ error. 
 The dashed line shows the optical depth relation for 
 the full pixel sample after masking out regions within $\pm170$~\kmps\ of known 
 galaxy redshifts, where we have considered galaxies with impact parameters of up to 2~pMpc.
 For a given \hone\ optical depth,
 we find a strong enhancement for the median \osix\ optical depth for pixels at small
 galactocentric distances. \textit{Centre and right panels:} The same as the left
 panel except for \osix(\cfour) and \osix(\sifour).
 The fact that we observe the same trend when binning by different
 ions along the x-axis provides additional evidence that the effect we are seeing is real. 
}
 \label{fig:o6_vs_h1_a}
\end{figure*}

\begin{table}
\caption{The median pixel optical depth values plotted in Figure~\ref{fig:o6_vs_h1_a}.}
\label{tab:ratios}
\begin{tabular}{@{}rrrrrr}
\hline\hline
\multicolumn{2}{l}{All pixels}& \multicolumn{2}{l}{0.04--0.18 pMpc}\\
& & \multicolumn{2}{l}{$\Delta v = \pm 170$~\kmps}\\
\hline
$\log_{10} \tau_{\honem}$ & $\log_{10} \tau_{\osixm}$ & $\log_{10} \tau_{\honem}$ & $\log_{10} \tau_{\osixm}$\\ 
$-1.94$ & $-1.73_{-0.02}^{+0.02}$ & $-2.01$ & $-1.91_{-4.09}^{+0.24}$ \\ 
$-1.49$ & $-1.72_{-0.02}^{+0.02}$ & $-1.45$ & $-1.01_{-0.21}^{+0.27}$ \\ 
$-1.02$ & $-1.70_{-0.02}^{+0.02}$ & $-0.95$ & $-1.07_{-0.16}^{+0.21}$ \\ 
$-0.53$ & $-1.65_{-0.03}^{+0.02}$ & $-0.53$ & $-1.28_{-0.16}^{+0.13}$ \\ 
$-0.04$ & $-1.58_{-0.03}^{+0.03}$ & $0.04$ & $-0.95_{-0.24}^{+0.14}$ \\ 
$0.44$ & $-1.33_{-0.03}^{+0.03}$ & $0.48$ & $-0.77_{-0.20}^{+0.12}$ \\ 
$0.96$ & $-1.07_{-0.04}^{+0.04}$ & $1.02$ & $-0.68_{-0.10}^{+0.09}$ \\ 
$1.47$ & $-0.80_{-0.06}^{+0.05}$ & $1.47$ & $-0.43_{-0.21}^{+0.09}$ \\ 
$1.96$ & $-0.78_{-0.08}^{+0.07}$ & &  \\ 
$2.48$ & $-0.66_{-0.17}^{+0.15}$ & &  \\ 
\hline
$\log_{10} \tau_{\cfourm}$ & $\log_{10} \tau_{\osixm}$ & $\log_{10} \tau_{\cfourm}$ & $\log_{10} \tau_{\osixm}$\\ 
$-2.47$ & $-1.69_{-0.02}^{+0.02}$ & $-2.48$ & $-1.30_{-0.15}^{+0.10}$ \\ 
$-2.01$ & $-1.64_{-0.02}^{+0.02}$ & $-2.00$ & $-1.18_{-0.14}^{+0.08}$ \\ 
$-1.61$ & $-1.46_{-0.03}^{+0.03}$ & $-1.57$ & $-0.82_{-0.10}^{+0.06}$ \\ 
$-1.05$ & $-0.77_{-0.05}^{+0.05}$ & $-1.03$ & $-0.56_{-0.13}^{+0.11}$ \\ 
$-0.55$ & $-0.35_{-0.05}^{+0.07}$ & $-0.57$ & $-0.33_{-0.19}^{+0.15}$ \\ 
$-0.05$ & $-0.13_{-0.08}^{+0.05}$ & $0.08$ & $0.05_{-0.06}^{+0.03}$ \\ 
$0.38$ & $0.12_{-0.07}^{+0.05}$ & &  \\ 
\hline
$\log_{10} \tau_{\sifourm}$ & $\log_{10} \tau_{\osixm}$ & $\log_{10} \tau_{\sifourm}$ & $\log_{10} \tau_{\osixm}$\\ 
$-2.45$ & $-1.69_{-0.02}^{+0.02}$ & $-2.44$ & $-0.83_{-0.13}^{+0.07}$ \\ 
$-2.01$ & $-1.65_{-0.02}^{+0.02}$ & $-2.02$ & $-1.15_{-0.16}^{+0.19}$ \\ 
$-1.61$ & $-1.60_{-0.03}^{+0.03}$ & $-1.55$ & $-0.91_{-0.13}^{+0.10}$ \\ 
$-1.09$ & $-1.50_{-0.06}^{+0.07}$ & $-1.04$ & $-0.82_{-0.22}^{+0.16}$ \\ 
$-0.54$ & $-0.66_{-0.13}^{+0.10}$ & $-0.47$ & $0.02_{-0.04}^{+0.03}$ \\ 
$-0.07$ & $-0.24_{-0.14}^{+0.11}$ & $-0.06$ & $0.10_{-0.01}^{+0.01}$ \\ 
$0.36$ & $-0.21_{-0.40}^{+0.28}$ & &  \\ 
\hline
\end{tabular}
\end{table}

The derived \osix(\hone) relations are shown in the left panel of Fig.~\ref{fig:o6_vs_h1_a},
where the black points and red stars indicate the relations for the full pixel
sample and the small galactocentric distance sample, respectively.
We also provide the data values shown in this figure in Table~\ref{tab:ratios}. 
The behaviour of the \osix(\hone) relation for the full pixel sample 
is consistent with previous measurements \citep[e.g.][]{schaye00a, aguirre08}
and we briefly explain the observed characteristics here. 

Focusing on the black circles, there are two distinct regions in \hone\ optical depth,
separated by\footnote{As noted in 
\citet{aguirre08}, because the \osix\ pixel relations are not as strong as for \cfour,
we fix the value of $\tau_{\rm cut}$ by hand rather than using functional fits.
We use the same values as in \citet{aguirre08} of $\tau_{\rm cut}$ = 0.1 
when pixel pairs are binned based on the optical depths of transitions 
that fall blueward of the QSO's \lya\ emission (\osix, \hone) and 
0.01 for those falling redward (\cfour, \sifour).}  $\tau_{\honem, {\rm cut}}\sim0.1$.
For $\tau_{\honem}>\tau_{\honem, {\rm cut}}$, $\tau_{\osixm}$ increases
with $\tau_{\honem}$. This relation arises because a large number of these pixel 
pairs are probing
regions which have been enriched by oxygen, and the value at each pixel
is set by the median number density ratio of \osix\ to \hone. 

For lower values of $\tau_{\honem}$ (below $\tau_{\honem,{\rm cut}}$),
$\tau_{\osixm}$ stays approximately constant, which indicates that the measured value is 
determined by residual contamination or noise and that the true, median value of 
$\tau_{\osixm}$ is below this detection limit. In general, this asymptotic value of $\tau_{\osixm}$ 
is slightly less than the median \osix\ optical depth of the full sample of pixels 
(the value is given in Table~\ref{tab:taumed}, and is indicated 
by the horizontal dotted black line in Fig.~\ref{fig:o6_vs_h1_a}).  
We attempt to measure the constant value to which the \osix\ pixel optical depths
asymptote, which we call the \osix\ flat level.
As in \citet{aguirre08}, we take this flat level to be the median of all $\tau_{\osixm}$ pixels
associated with $\tau_{\honem} < \tau_{\honem, {\rm cut}}$.  

Furthermore, we estimate a 1-$\sigma$ error on this
quantity by dividing each spectrum into 5~\AA\ chunks and creating 1000 bootstrap 
resampled spectra. The flat level and the associated 1$\sigma$ error are denoted by the black dashed line
and grey region, respectively. 
It is difficult to probe \osix\ optical depths below the flat level, as one becomes 
limited by contamination. Specifically, absorption from ions with rest wavelength less than that
of \hone\ \lya\ are found bluewards of the QSO \lya\ emission line, 
and their recovery is limited by contamination from \hone\ and other metal lines. 
These metal transitions are less affected by the quality (S/N) of the spectra and are
more sensitive to the method of recovery. On the other hand, for ions redwards  
of the QSO's \lya\ emission, the median pixel optical depths are set mainly by 
the S/N and/or shot-noise, since the majority of pixels do not have detectable 
metal absorption. The two right panels of Figs.~4 and~5 in \citet{aguirre02}
show the changes on the median levels for different recovery methods and S/N ratios for \osix\
(bluewards of the QSO \lya\ emission) and \cfour\ (redwards of the QSO \lya\ emission)
using simulated spectra. These figures clearly demonstrate how the median level for \osix\ is more
sensitive to recovery method while for \cfour\ changing the S/N has a greater effect.

Turning next to the red stars in the left panel of Fig.~\ref{fig:o6_vs_h1_a}, 
we see a significant enhancement of $\tau_{\osixm}$ at
fixed $\tau_{\honem}$ for the small galactocentric distance pixels 
compared to the full pixel sample. 
Such a difference is not present if we consider larger impact parameter 
bins (not shown). We emphasize that although the full pixel sample
is representative of random regions, it
is composed of pixels both near and far from galaxies (and many of these galaxies
are likely not detected in our survey). To demonstrate this, 
we examine the effect of masking out the regions in the spectra
that are known to be near galaxies. We consider all 
galaxies in our sample with impact parameters $\leq2$~pMpc and with
redshifts satisfying eq.~\ref{eq:zlim}, and mask out regions of $\pm170$~\kmps\ around
these redshifts in all 15 of our QSO spectra. The resulting optical depth relation
is shown as the dashed line in the left panel of Fig.~\ref{fig:o6_vs_h1_a}, 
and we conclude that the full pixel sample relation is largely independent of the
presence of regions proximate to detected galaxies.

Next, we vary the ion plotted along the x-axis. 
In the centre and right panels of Fig.~\ref{fig:o6_vs_h1_a}, we
show the relations for \osix(\cfour) and \osix(\sifour), respectively 
(the values are also given in Table~\ref{tab:ratios}). 
Just as for \osix(\hone), we observe significantly enhanced \osix\ optical depths 
at both fixed $\tau_{\cfourm}$ and fixed $\tau_{\sifourm}$.
The persistent enhancement of OVI absorption at fixed optical depth of three 
distinct ions (\hone, \cfour\ and \sifour) makes the individual detections with respect 
to each ion still more significant. 
Here we note that although the \osix(\hone) and \osix(\cfour) optical depth 
relations for which the galaxy positions have been masked out of the spectra
show significant signal, this is not the case for \osix(\sifour). 
This suggests that strong \sifour\ absorption arises primarily
near galaxies.

\begin{figure*}
 \includegraphics[width=\wc]{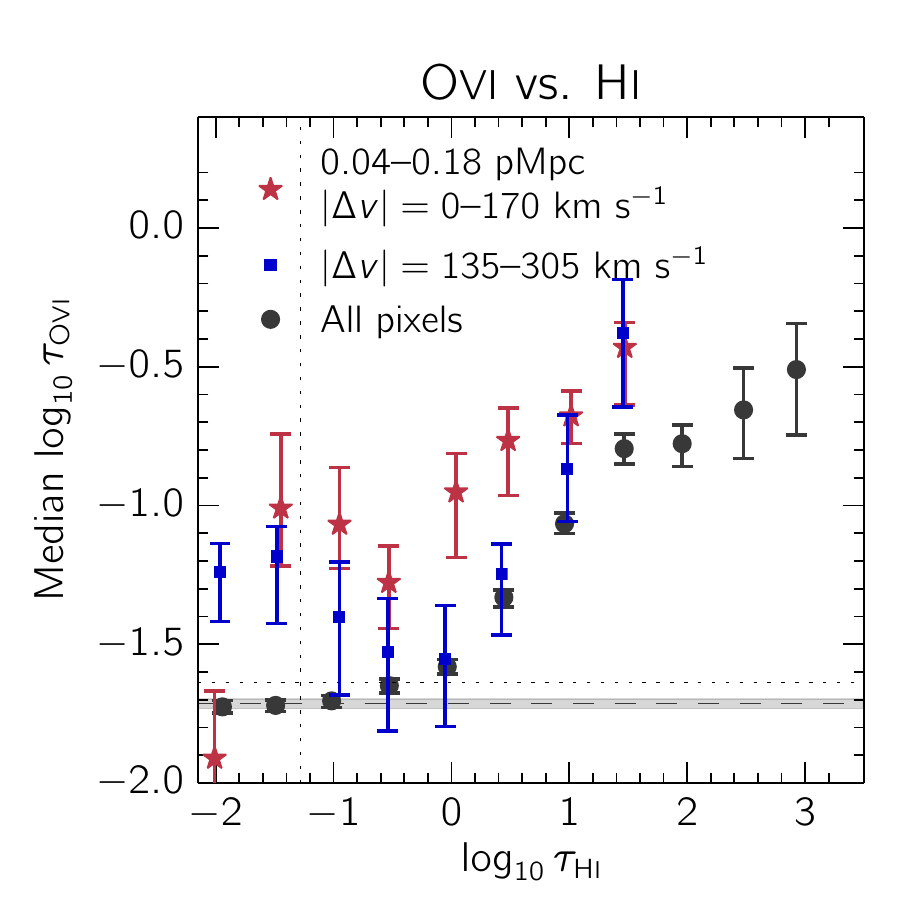} 
  \includegraphics[width=\wc]{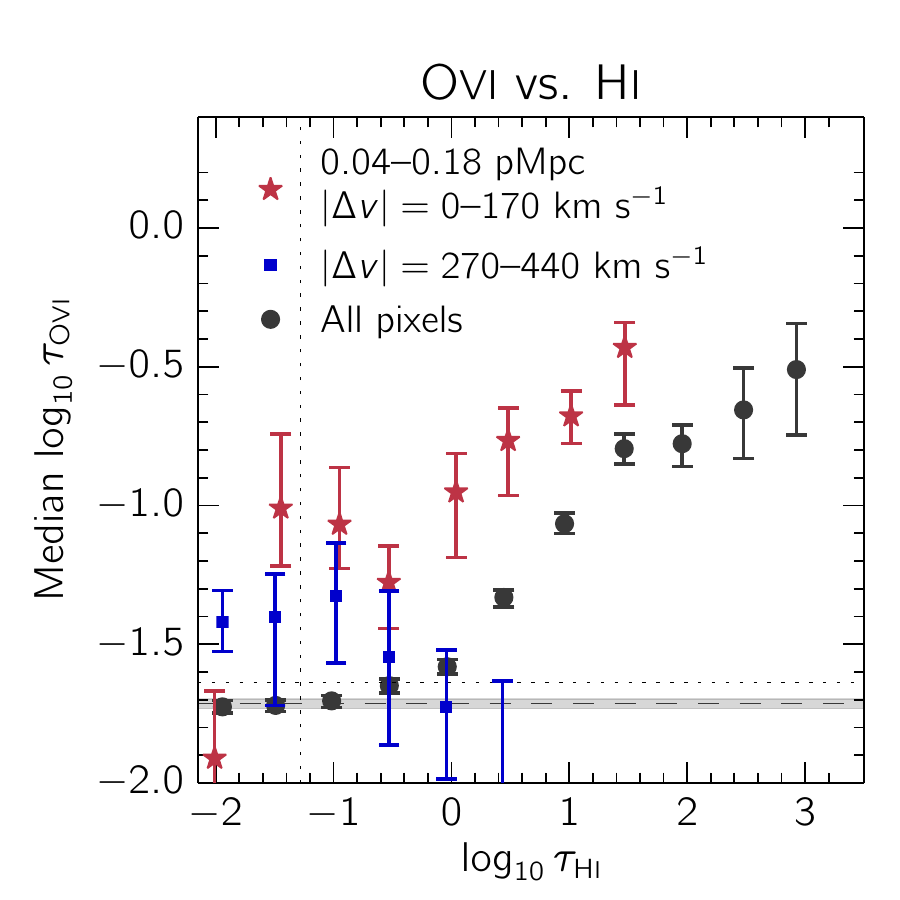} 
 \includegraphics[width=\wc]{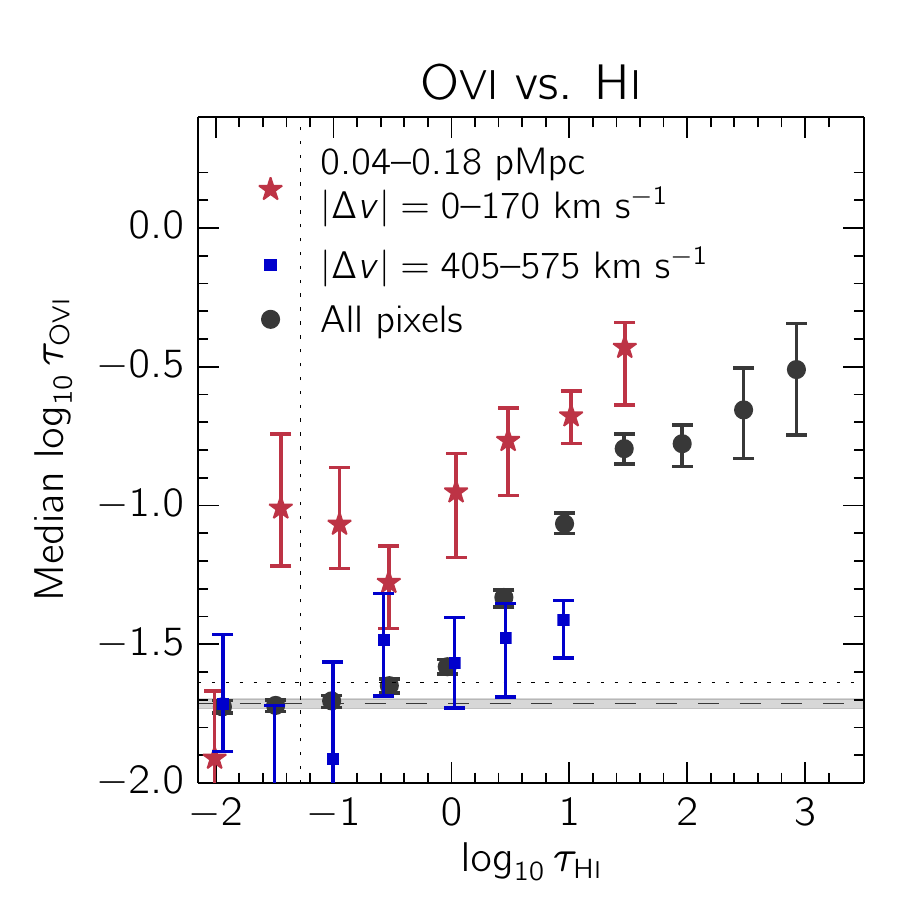} \\
 \caption{The same as Fig.~\ref{fig:o6_vs_h1_a} but we have overplotted the 
  result of taking different velocity cuts when considering the small galactocentric
  distance points (blue squares). We have chosen velocity bins such that 
  each cut spans the same velocity range of 340~\kmps, and find that
 the enhancement of \osix\ at fixed $\tau_{\honem}$ is present for 
 $\tau_{\honem}\lesssim0.1$ out to the $|\Delta v| = 270\text{--}440$~\kmps\ velocity cut.
 Using the central value of this velocity range as our upper limit, we conclude that the enhancement persists
out to $\sim350$~\kmps, which is $\gtrsim1.5$ times the typical circular 
  velocities of the galaxies in our sample. 
}
 \label{fig:vary_vel}
\end{figure*}

It is instructive to assess how sensitive the enhancement is
to the chosen velocity range. 
We have examined
the \osix(\hone) relation for velocity bins starting with $|\Delta v| = 0\text{--}170$~\kmps\ and
increasing both velocity limits by increments of 10~\kmps
(such that each cut spanned the same total velocity range of 340~\kmps). We 
 found that for optical depth bins with $\tau_{\honem}\lesssim1$, the enhancement
is present up to a velocity range of $|\Delta v| = 270\text{--}440$~\kmps, and we
take the midpoint of this bin, $\sim350$~\kmps, as an upper limit to the 
extent of the \osix\ enhancement. 
In Fig.~\ref{fig:vary_vel} we show the relation for the  $|\Delta v| = 270\text{--}440$~\kmps\ velocity
bin (central panel), as well as higher and lower velocity cuts (right and left panels, respectively). 
For most of the optical depth bins with $\tau_{\honem}>1$,
the enhancement in \osix\ is only significant for velocities 
within $\pm170$~\kmps\ of the galaxy positions. 

Although for the remainder of the analysis we will continue using the smallest velocity
cut of $|\Delta v| =0\text{--}170$~\kmps\, we emphasize that for the lowest \hone\ optical 
depth bins, \osix\ is enhanced out to velocities of $\sim350$~\kmps, corresponding to 
$\gtrsim1.5$ times the typical circular velocities of the galaxies in our sample. 
While simulations predict that absorbers from galaxies below the detection limit 
can be projected to such velocities around their more massive counterparts \citep[e.g.,][]{rahmati14},
in \S~\ref{sec:models} we determine that only a hot, collisionally ionized gas phase
is observed out to these large velocities, which is certainly suggestive of outflows. 
Further comparisons with simulations
will be required to fully disentangle outflow and clustering effects.

\subsection{Are the observed differences in optical depth ratios real?}
\label{sec:testing}

\begin{figure*}
  \includegraphics[width=\wc]{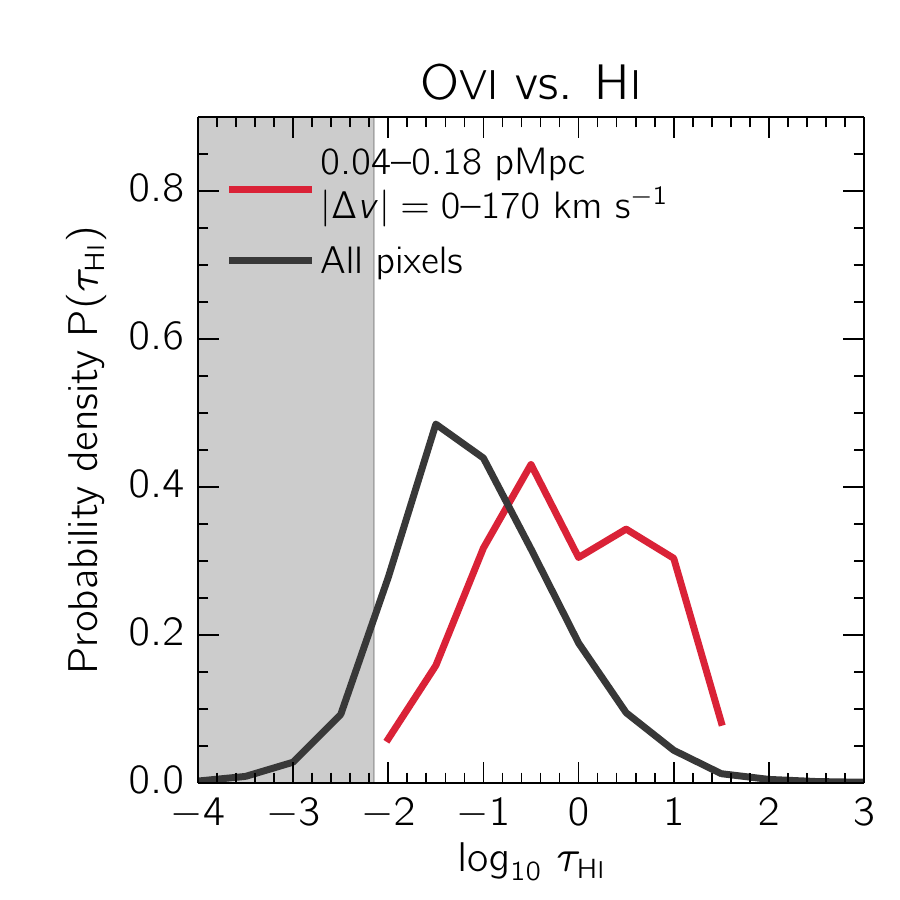} 
  \includegraphics[width=\wc]{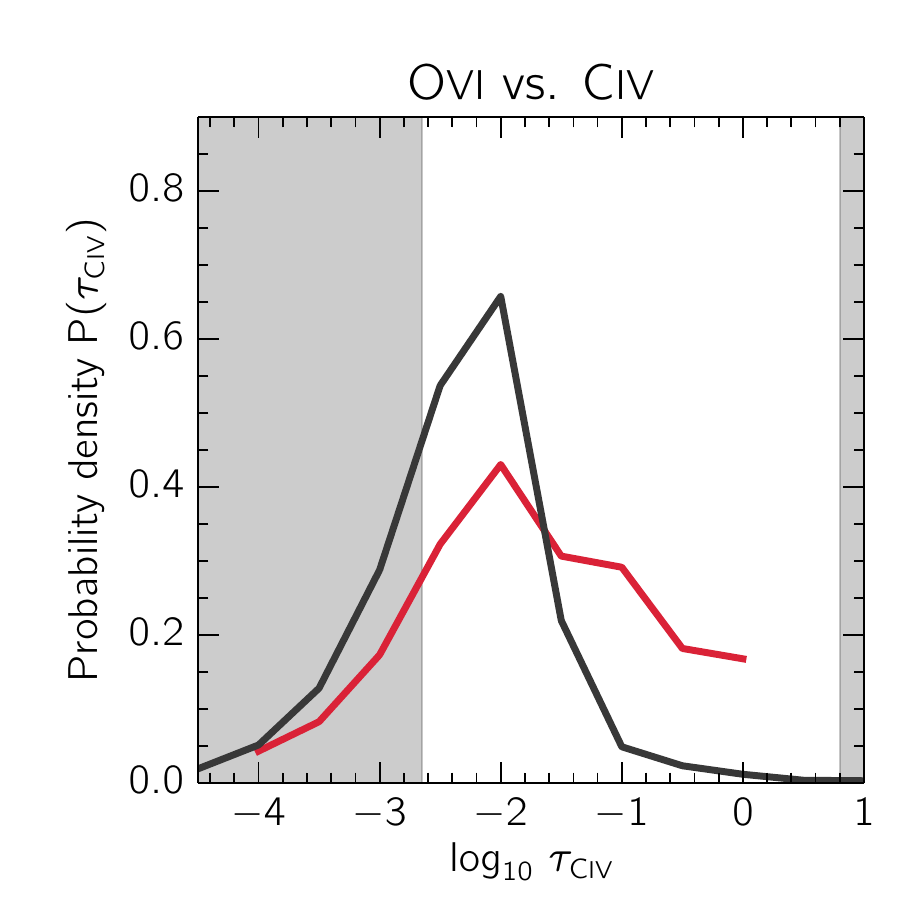} 
  \includegraphics[width=\wc]{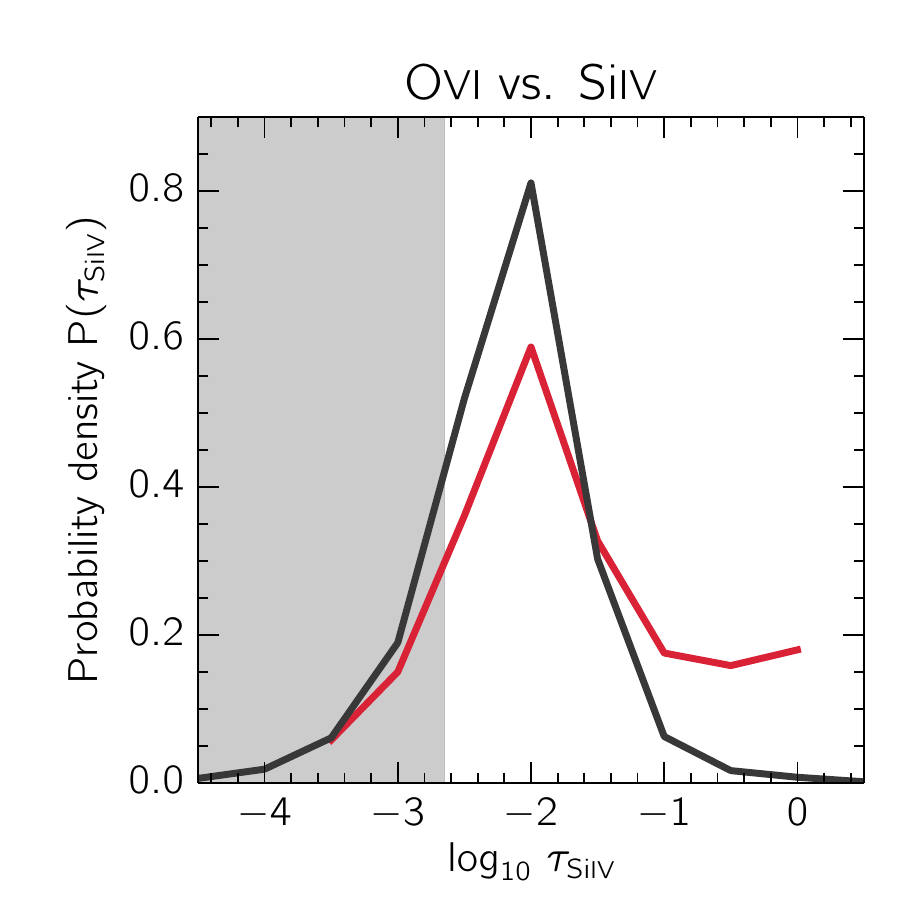} 
  \caption{The probability density function of \hone\ (left), \cfour\ (centre)  
	and \sifour\ (right) pixel optical depths for both the small galactocentric distance sample
        (red) and the full sample (black). For ease of comparison,
         we have shaded regions along the x-axis that are not included in the previous figures.} 
 \label{fig:odhist}
\end{figure*}

\begin{figure*}
  \includegraphics[width=\wc]{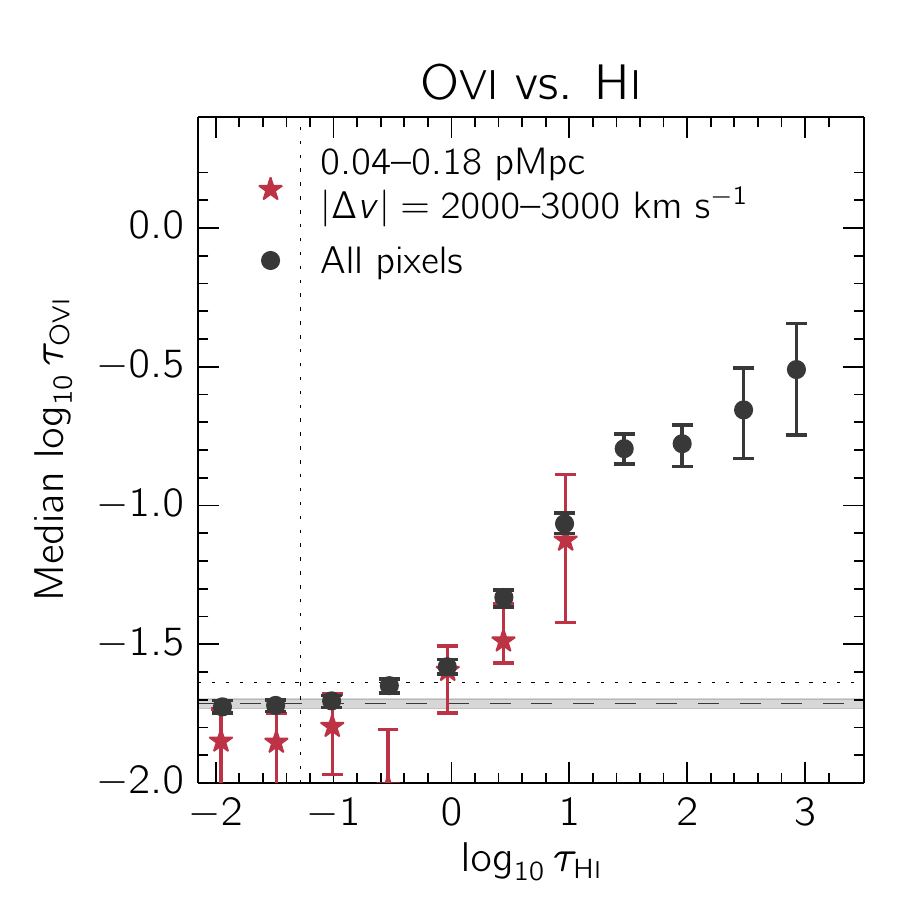} 
  \includegraphics[width=\wc]{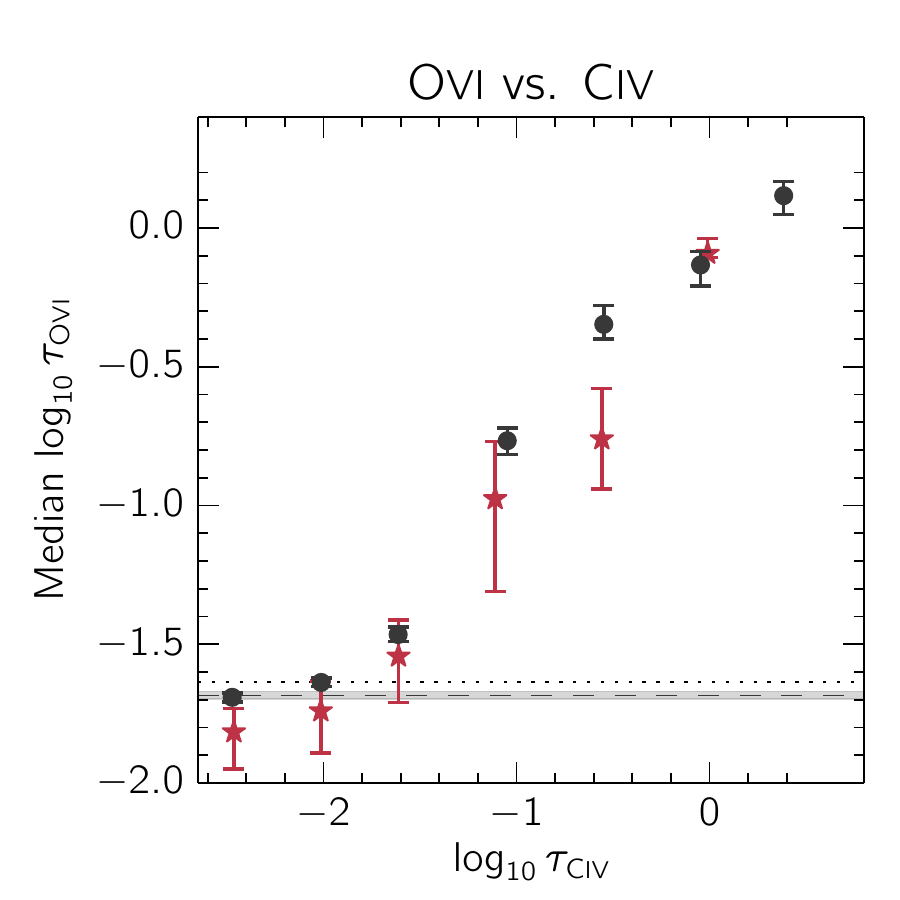} 
  \includegraphics[width=\wc]{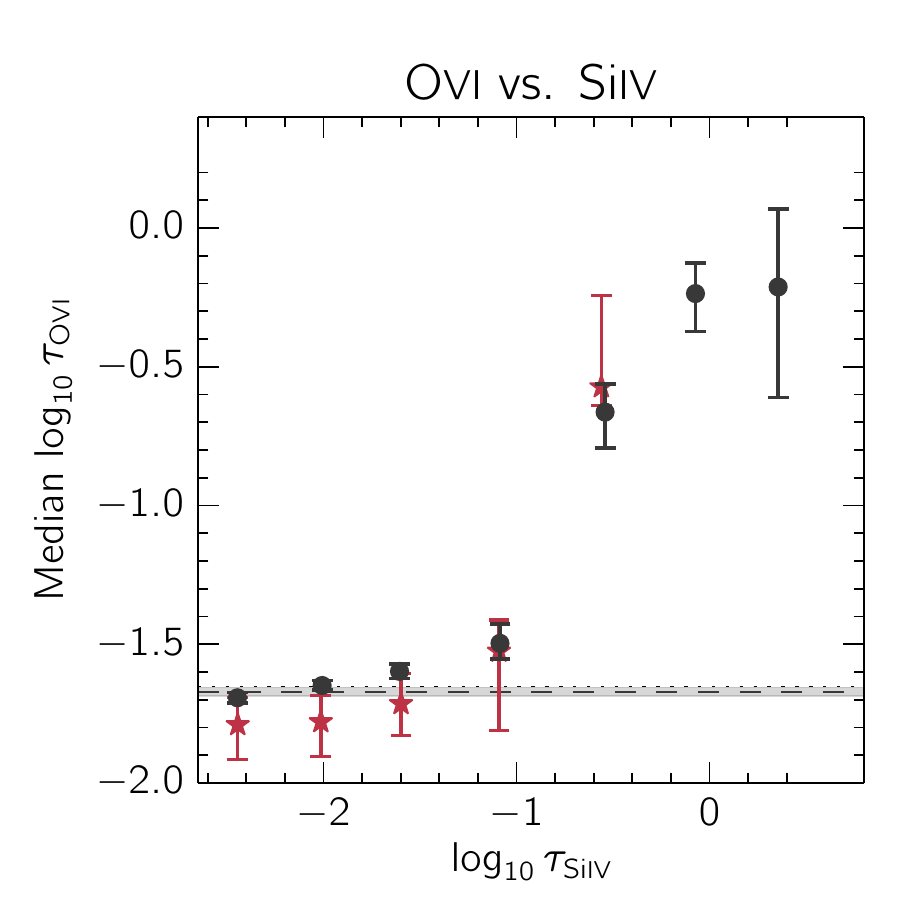} \\
\caption{The same as Fig.~\ref{fig:o6_vs_h1_a} 
  but instead of taking pixels within $\pm170$~\kmps
  of the galaxy redshifts,
  the red stars show the effect of choosing regions within
   $|\Delta v| = 2000\text{--}3000$~\kmps\ of the galaxy redshifts. By excluding the positions directly
   around the galaxies, but still using pixels within $\sim10^3$~\kmps\ of the galaxy
   redshifts, we remove physical effects caused by the presence of the galaxy while
    probing the same spectral properties such as S/N, resolution, and distance
    from the QSO. Contrary to Fig.~\ref{fig:o6_vs_h1_a}, the \osix\ absorption 
is not enhanced for the small galactocentric distance sample, which implies that 
the enhancement visible in that figure is not due to systematic differences in the spectral properties between the two samples.
}
 \label{fig:outer_vel}
\end{figure*}

\begin{figure*}
\includegraphics[width=\wc]{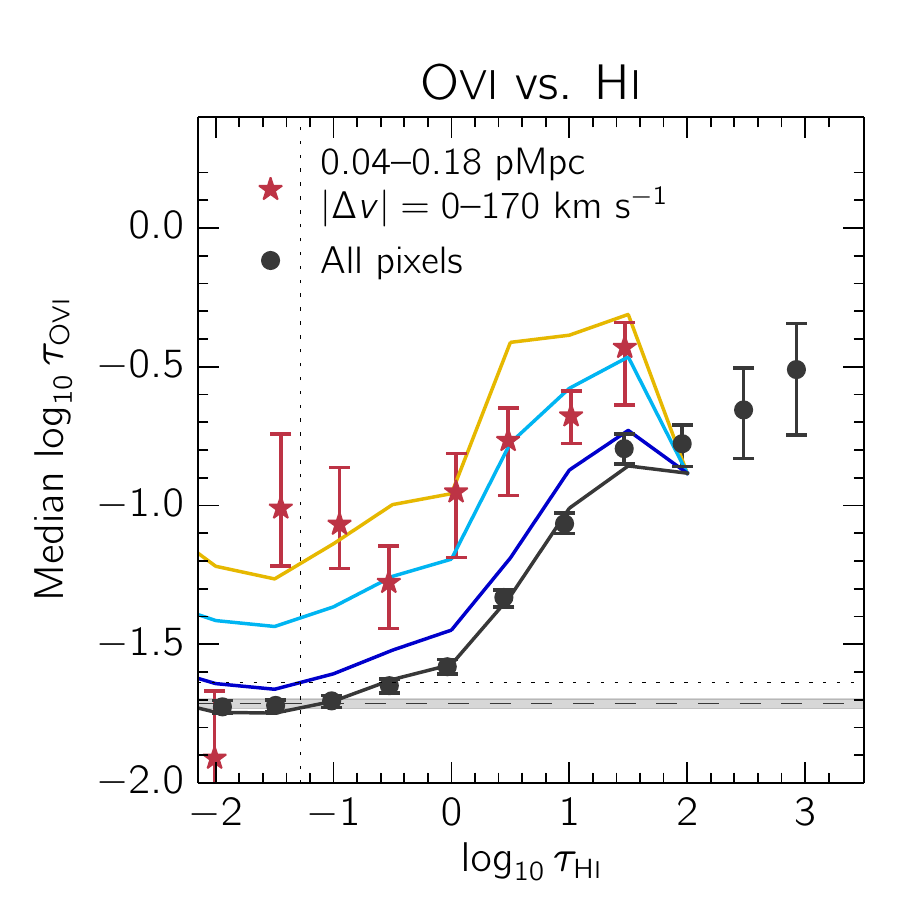} 
\includegraphics[width=\wc]{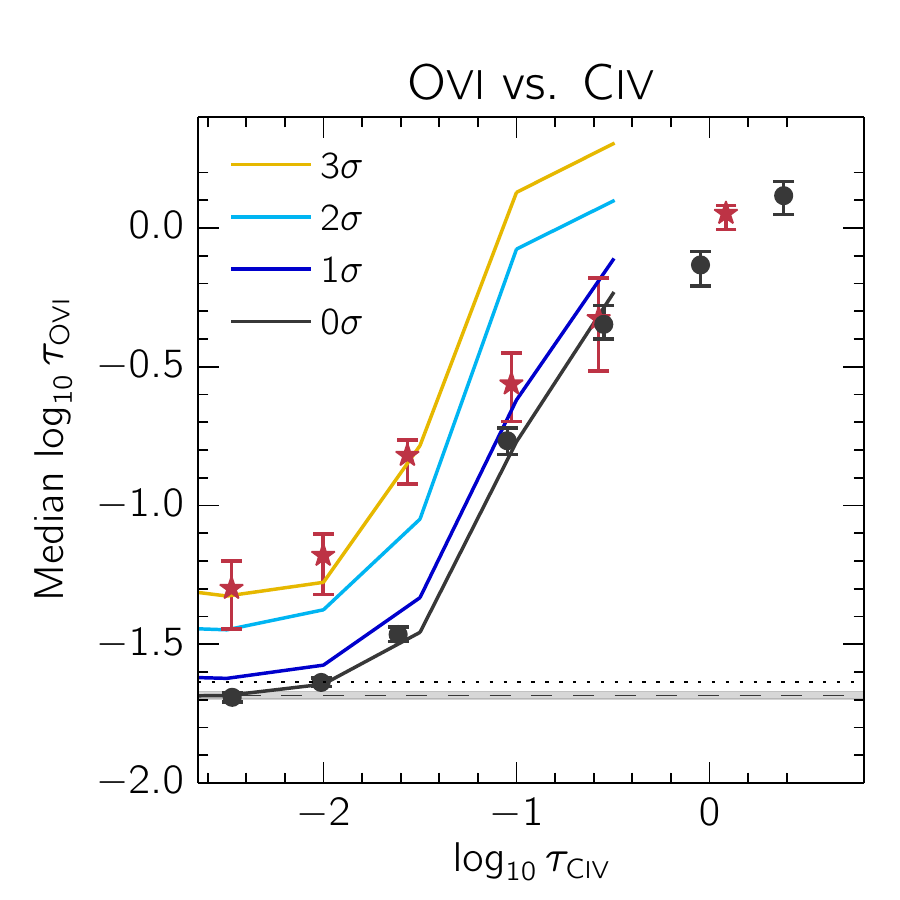} 
\includegraphics[width=\wc]{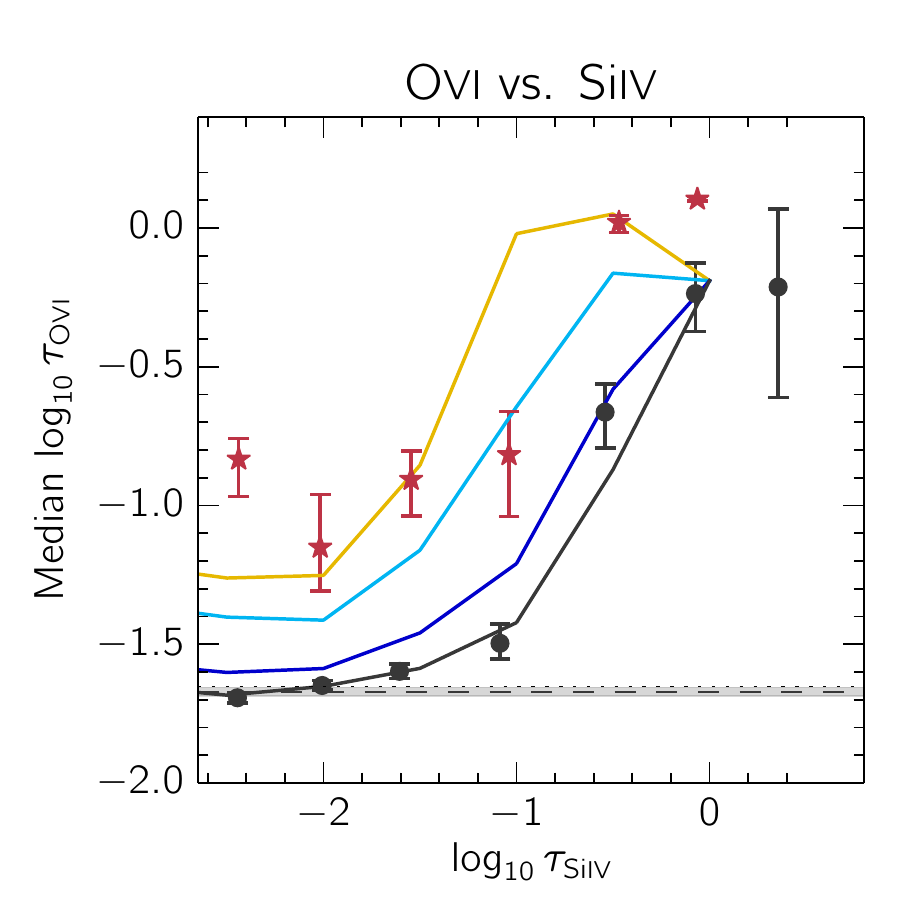} 
\caption{The same as Fig.~\ref{fig:o6_vs_h1_a}, except that we have overplotted
the results of randomising the redshifts of the small
galactocentric distance galaxies 1000 times. For each of these randomizations,
we recompute \osix(\hone), \osix(\cfour) and \osix(\sifour), and in this figure we show the median, 
and the 84.1, 98.7, and 99.8 percentiles (corresponding to 0, 1, 2, and 3$\sigma$, 
respectively) from the full distribution of realizations. 
We find that the enhancement of \osix\ for fixed $\tau_{\osixm}$,
$\tau_{\cfourm}$ and $\tau_{\sifourm}$ is approximately a 2--3-$\sigma$ effect per bin.}
\label{fig:random_reg}
\end{figure*}

Fig.~\ref{fig:o6_vs_h1_a} suggests that the gas near the sample galaxies has properties different 
from that in random regions with the same strength of \hone, \cfour, or \sifour\ optical depth. 
However, it is important to verify that the difference
between the two pixel samples is
not driven primarily by chance or systematic errors. In 
particular, limiting ourselves to only a few regions of the spectra could skew the results
if, for example, these regions have different S/N, or inconsistent contamination
levels due to being located at different redshifts. 

First, we would like to be sure that the enhancement in the median \osix\ optical 
depths is not due to a small number of pixels. In Fig.~\ref{fig:odhist}
we show the pixel optical depth probability density functions (PDFs)
for \hone, \cfour\ and \sifour, for both the small galactocentric distance
and the full pixel samples (note that the x-axis ranges are larger than those 
shown in Fig.~\ref{fig:o6_vs_h1_a}). Focusing on the small galactocentric 
distance pixels (red lines), one sees that the bins with enhanced median
\osix\ ($\log_{10} \tau_{\honem} \gtrsim -1$;
$\log_{10} \tau_{\cfourm}$ and $\log_{10} \tau_{\sifourm} \gtrsim -2$) comprise 
 the majority of the pixels. Thus, the enhancement
in the median \osix\ optical depth cannot be attributed to a small number of pixels. 

As a further test, in Fig.~\ref{fig:outer_vel} we plot $\tau_{\osixm}$
versus $\tau_{\rm x}$ for the same $\tau_{\rm x}$ shown in Fig.~\ref{fig:o6_vs_h1_a}, but instead of using pixels
within $\Delta v = \pm170$~\kmps\ of the galaxy redshifts, we look at regions
further away, i.e.\ $ | \Delta v| =  2000 \mbox{--}3000$~\kmps. This velocity cut is near
enough to the galaxy redshifts that 
within each impact parameter bin we are still looking at areas of the spectra with the 
same S/N and contamination characteristics, but far enough to avoid regions that may
be associated with the galaxies. 
If the enhancement of the \osix\ optical 
depth at fixed \hone, \cfour, or \sifour\ that we detect near galaxies were caused by systematic differences in the spectral
properties of the two samples, then we would expect to see the enhancement to a similar 
significance in these figures. However, in every case where we previously 
saw an enhancement in the median \osix\ optical depth for the pixels at small 
galactocentric distance, the effect is completely removed in Fig.~\ref{fig:outer_vel}. 

Next, we examine whether the optical depth differences
are consistent with random fluctuations. To do this, we take the galaxies
from the small galactocentric distance sample, randomize their redshifts 1000 times,
and calculate \osix(\hone), \osix(\cfour) and \osix(\sifour)
for every random realization. In Fig.~\ref{fig:random_reg}
we show the one-, two-, and three-$\sigma$ percentiles that result 
from this procedure. From this, we measure the enhancements seen
in \osix(\hone), \osix(\cfour), and \osix(\sifour)
to be approximately a 2--3-$\sigma$ effect per bin. 

Additionally, because we do not observe differences in   
the \osix(\hone), \osix(\cfour) and \osix(\sifour) relations for 
galaxies with impact parameters $>180$~pkpc and the full pixel sample,
here we check whether the \osix\ enhancement for the galaxies with 
impact parameters $<180$~pkpc can be attributed to properties 
other than the galaxy distance to the QSO sightline. 
We directly test whether the following three characteristics
in every impact parameter bins are consistent with the full sample:
(1) the galaxy redshifts (lower galaxy redshifts mean that \osix\ is 
more contaminated by \hone), (2) the galaxy velocity distance from the QSO \lya\ emission
(to rule out QSO proximity effects) and (3) the S/N of the spectral regions
(smaller S/N will bias the optical depth estimation high).  

We have measured the $p$-values resulting
from a 2-sample Kolmogorov-Smirnoff test, comparing the galaxy 
redshift, median S/N within $\pm170$~\kmps\ from the redshift of the galaxy, and redshift difference
between the galaxy and the QSO, between the galaxies in the 
small galactocentric distance sample and the full galaxy sample. In every instance
we find $p$-values greater than 0.1, which is consistent with 
the null hypothesis that the two samples are drawn from the same distribution.

Finally, in Appendix~\ref{sec:o6_test} we have tested how the \osix\ contamination
correction affects our results. We determined that although the inferred values of the \osix\ optical depths
are sensitive to changes in the correction procedure, the enhancement of \osix\ optical depths at fixed 
\hone\ the \osix\ for pixels at small galactocentric distances compared to random regions is unchanged,
and remains significant irrespective of the contamination correction details. 
Thus, we conclude that the observed optical depth differences  
are neither due to chance nor to systematic variations in the spectral characteristics
in either the QSOs or the galaxies.

\begin{figure}
 \includegraphics[width=0.45\textwidth, trim=0mm 2mm 0mm 0mm]{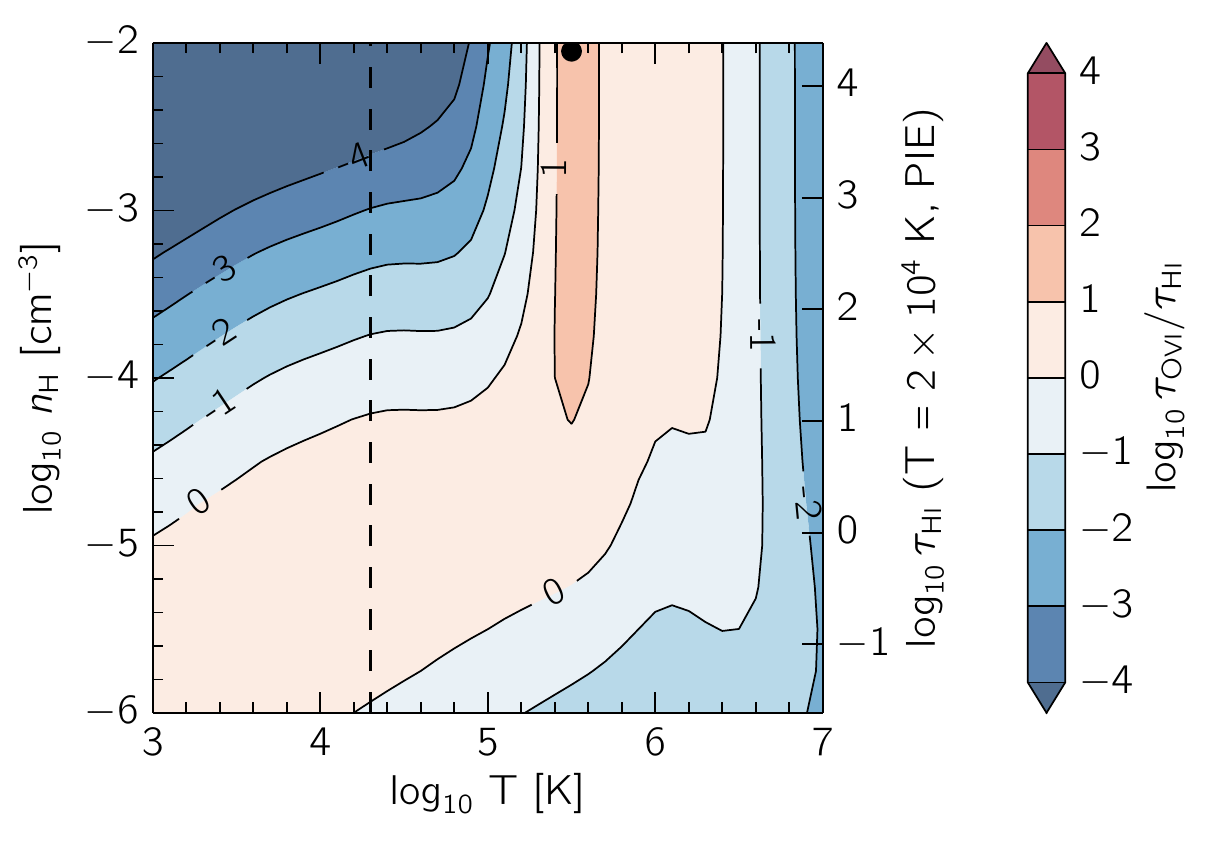}\\
 \caption{Theoretical optical depth ratios ($\log_{10} \tau_{\osixm} / \tau_{\honem}$)
as a function of temperature and hydrogen density from \texttt{CLOUDY} modelling. For this figure
 we have assumed $\text{[O/H]}=0.0$, however we allow this value to vary according to eq.~\ref{eq:metallicity}
when comparing with the observations. The vertical dashed line denotes $T=2\times10^4$~K, which 
is the temperature that we assume in the case of PIE. Furthermore, under the assumption of PIE 
we can convert hydrogen number densities to \hone\ optical depths with eq.~\ref{eq:nh}, 
and we show corresponding $\tau_{\honem}$ values on the right-hand y-axis. 
For CIE, we set can use the ratios to set lower limits on the metallicity by using the maximum
ratio from this temperature-density plane. This maximum occurs at $T=3\times10^5$~K and is marked 
by the small black circle at $n_{\rm H}=-2$. 
}
 \label{fig:cloudy_a}
\end{figure}

\section{Ionization models}
\label{sec:models}

In this section, we will investigate the physical origin of the 
difference in \osix\ observed between regions near our galaxies and in random locations. 
We focus on \osix(\hone) because binning by $\tau_{\honem}$ enables a more straightforward 
physical interpretation than binning by metal ion optical depth, due to the tight correlation 
between \hone\ absorption and gas density in photoionized gas. We consider three 
possible scenarios and examine their plausibilities. The first hypothesis 
we consider is enriched photoionized ($T\sim10^4$~K) gas, where 
the enhanced \osix(\hone) near galaxies might be explained by higher gas metallicities. Second, 
we test the idea that ionizing radiation from the nearby galaxy
could be responsible for the increase in \osix(\hone). 
Finally, 
we consider whether our observations can be explained by a hot, 
collisionally ionized enriched gas phase near galaxies. 
We argue that of these three, the first scenario (enriched photoionized gas)
can account for pixels with $\tau_{\honem}\gtrsim10$, while 
only the third explanation (the presence of hot, enriched gas) 
is plausible for pixels with $\tau_{\honem}\lesssim1$. 
Of course, it is important to note that every
\hone\ bin likely contains a mixture of pixels from different gas phases and ionization sources; 
the final behaviour is simply determined by the dominant phase.

\subsection{Photoionization by the background radiation}
\label{sec:pie}

The optical depth of \hone\ is believed to be a good tracer of the 
photoionized gas density,
even on an individual pixel basis \citep{aguirre02}.
Hence, if the gas probed were predominantly photoionized and if the abundance of 
oxygen depended only on gas density, 
we would not expect to see any difference
between the \osix(\hone) relations of all pixels 
and those known to lie near galaxies in Fig.~\ref{fig:o6_vs_h1_a}. 
Since a clear difference is observed, we postulate that this could be caused 
by an increase in the oxygen abundance near galaxies at fixed gas densities.

To test this idea, we turn to ionization modelling using \texttt{CLOUDY} \citep[version 13.03]{ferland13}.
Our setup involves a plane-parallel slab illuminated uniformly 
by an ionizing background, along a grid of varying temperatures 
and hydrogen densities, that covers both photoionization equilibrium
(PIE) and collisional ionization equilibrium (CIE, discussed in \S~\ref{sec:cie}). For the fiducial case we use the ionizing
background from \citet{haardt01} including contributions from both
quasars and galaxies, normalized to match the $z=2.34$ metagalactic 
\hone\ photoionization rate, $\Gamma = 0.74\times10^{-12}\,\text{s}^{-1}$, 
from \citet{becker07}.\footnote{ 
Measurements of $\Gamma$ at $z=2.4$ vary between $\Gamma=0.5 \times 10^{-12}\, \text{s}^{-1}$
in \citet{fauchergiguere08} up to
$\Gamma = 1.0 \pm^{0.40}_{0.26} \times 10^{-12}\,\text{s}^{-1}$ in
\citet{becker13}.
We choose the intermediate value of $0.74\times 10^{-12}\, \text{s}^{-1}$ at
$z=2.34$ taken from the fitting formulas of \citet{becker07}.
} However, the shape and normalization of the background is 
subject to large uncertainties, and we explore the effect of
varying them in Appendix~\ref{sec:bkgd}.\footnote{We note that the assumption of 
PIE  may break down as
 non-equilibrium effects become important in cooling
gas at temperatures $<10^6$~K. The presence of these effects causes 
collisional ionization to occur at lower temperatures compared to in equilibrium,
although once an extragalactic background
is included the impact of the effect of non-equilibrium 
cooling become less important \citep{oppenheimer13a}.}

\begin{table}
\caption{Solar abundances used in this work, taken from \texttt{CLOUDY}~13. 
 References are 1. \citet{allende02}; 2. \citet{allende01}; 3. \citet{holweger01}. }
\label{tab:abundances}
\centering
 \begin{tabular}{@{}ccc}
\hline
\hline
Element & $n_i/n_{\rm H}$ & Ref. \\
\hline
H 	&	$1$  			&	\\
C	&	$2.45\times10^{-4}$	&  1.	\\
O	&	$4.90\times10^{-4}$	&  2. \\
Si	&	$3.47\times10^{-5}$  	&  3. \\
\hline
\end{tabular}
\end{table}

After obtaining individual ionization fractions from the above models as a function of density, 
$n_{\rm H}$, and temperature, $T$, we can relate metallicity 
to optical depths using the following equation:
\begin{equation}
   [{\rm O/H}] = \log_{10} \dfrac{\tau_{\osixm}}{\tau_{\honem}} \dfrac{(f\lambda)_{\honem}}{(f\lambda)_{\osixm}}
    \dfrac{n_{\rm O}}{n_{\osixm}} \dfrac{n_{\honem}}{n_{\rm H}} - (\rm {O/H})_{\odot}
\label{eq:metallicity}
\end{equation}
where $f$ and $\lambda$ are the oscillator strengths and rest wavelengths of the transitions, 
and we assume solar abundances from \texttt{CLOUDY}~13 (listed in Table~\ref{tab:abundances}). The resulting 
optical depth ratio contours, as a function of temperature and hydrogen number 
density and assuming solar metallicity, are shown in Fig.~\ref{fig:cloudy_a}. 
At temperatures $T<10^5$~K, photoionization dominates and the contours are only weakly 
dependent on temperature, while at higher temperatures and for sufficiently high densities
collisional ionization dominates and the contours are independent of the density.

\begin{figure*}
 \includegraphics[width=\wc]{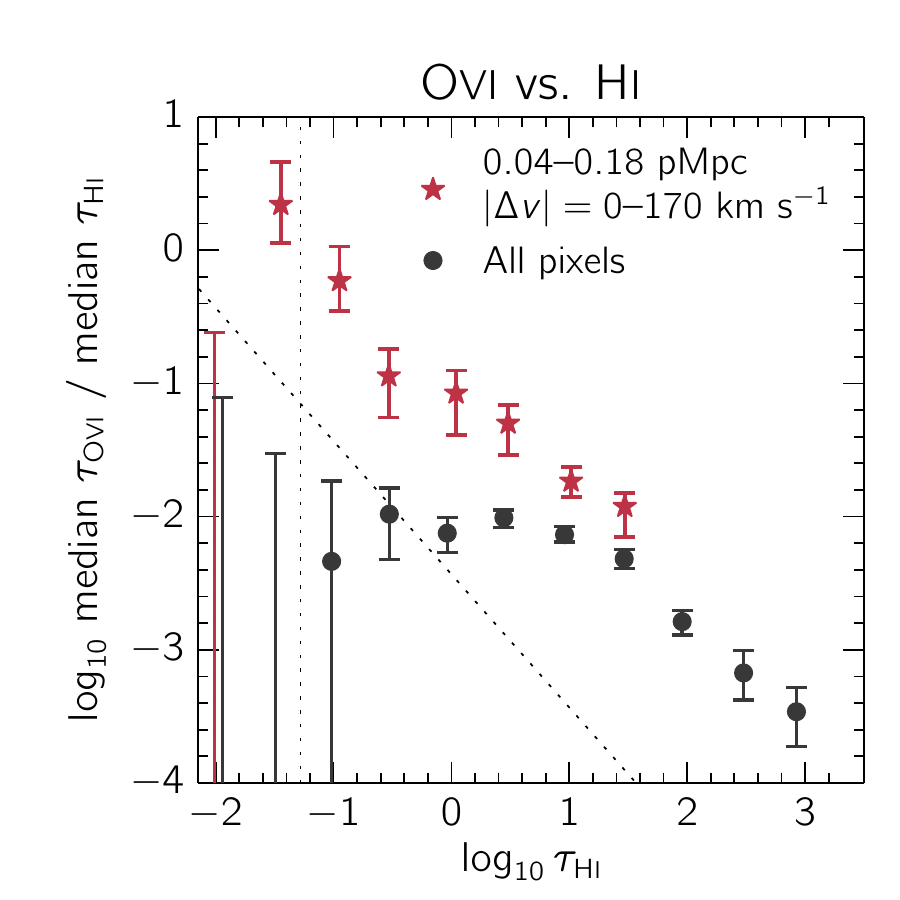}
 \includegraphics[width=\wc]{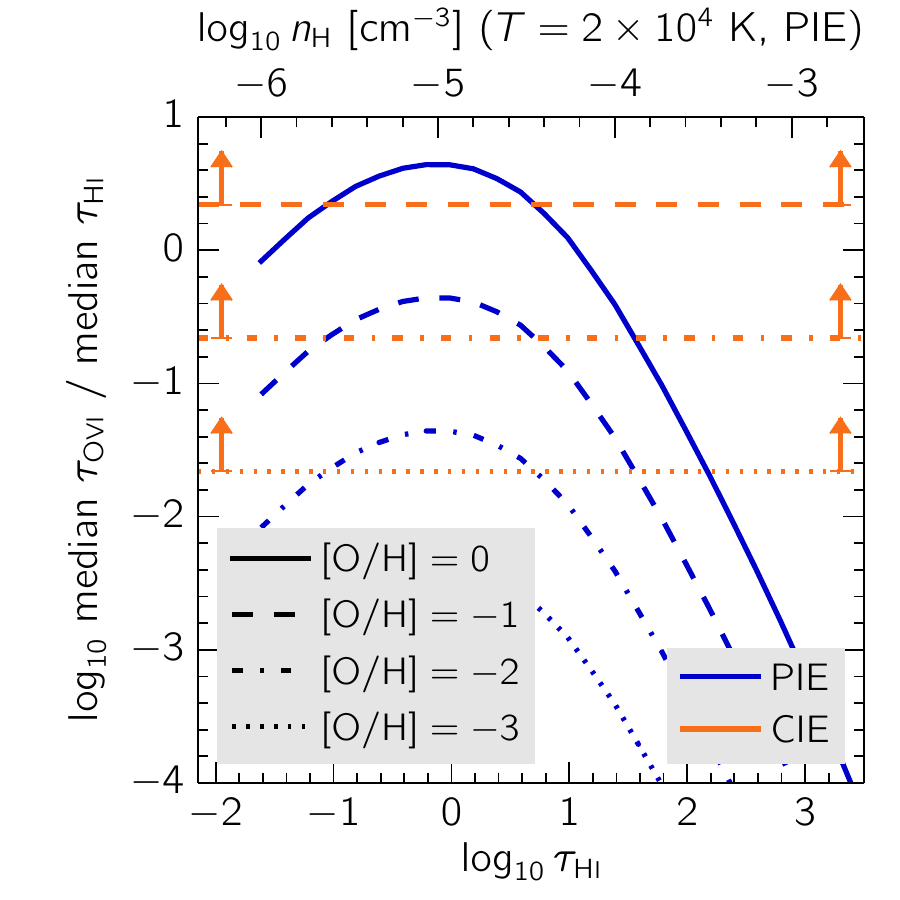}
 \caption{\textit{Left panel:} The ratio of \osix\ to \hone\ optical depths as a function
 of $\tau_{\honem}$. To compute these points, we used the data from
  Fig.~\ref{fig:o6_vs_h1_a}. First, we subtracted the \osix\ flat level from the
  median \osix\ optical depths, in order to correct for contamination.
  After dividing the resulting values by their corresponding \hone\ optical depths, we obtain
  the points plotted here.
  The black vertical dotted line corresponds
to the median value of all \hone\ pixel optical depths, while the diagonal line shows the median value of all
\osix\ pixel optical depths divided by the given value of $\tau_{\honem}$ along the x-axis.
\textit{Right panel:} Theoretical ratios of \osix\ to \hone\ as a function of $\tau_{\honem}$ taken 
 from \texttt{CLOUDY} modelling (see Fig~\ref{fig:cloudy_a}). First, in the 
 case of PIE,  we assume that the gas has a temperature of $2\times10^4$~K, and convert \hone\ optical
 depths to a hydrogen number density using eq.~\ref{eq:nh} (the corresponding $n_{\text{H}}$
values are shown on the upper y-axis). The blue curves show profiles taken along the vertical black
dashed line in Fig.~\ref{fig:cloudy_a}, where the different line styles demonstrate the effect of varying
the metallicity. We also consider CIE, in which case cannot 
estimate the density from $\tau_{\honem}$. Instead, we can use the maximum theoretical ratio
(indicated by the black circle at $T=3\times10^5$  Figure~\ref{fig:cloudy_a}) to set a lower 
limit on the metallicity, and these are shown by the horizontal orange lines. 
}
 \label{fig:o6_vs_h1_b}
\end{figure*}

We would like to compare the observed optical depth ratios with those predicted from \texttt{CLOUDY}
in order to estimate the gas metallicity.
To compute the observed ratios, we begin
with the $\tau_{\osixm}$ points in the left panel Fig.~\ref{fig:o6_vs_h1_a}. In order to
correct for the presence of residual contamination from absorption by species other than \osix, 
we subtract the 
flat level (i.e.\ the asymptotic value of $\tau_{\osixm}$ reached for small 
$\tau_{\honem}$ and indicated by the horizontal, dashed line in Fig.~\ref{fig:o6_vs_h1_a}) 
from all $\tau_{\osixm}$ values.
Following \citet{schaye03}, to be conservative we added the error on the flat level linearly 
(rather than in quadrature) to the errors on $\tau_{\osixm}$, and finally we divide every
$\tau_{\osixm}$ point by its corresponding $\tau_{\honem}$. The results of this calculation 
are plotted in the left panel Fig.~\ref{fig:o6_vs_h1_b}. 

Next, since we are assuming that the gas is photoionized, we can 
transform the \hone\ optical depths to gas densities, as was for example done in \citet{rakic12}. 
The relation between column density $N$ and pixel optical depth at the line centre, $\tau_0$, is:
\begin{equation}
\begin{split}
\label{eq:tau0}
\tau_{0} =& \dfrac{\sqrt{\pi}e^2}{m_e c} \dfrac{N\, f\, \lambda_0}{b_D} \\
  \approx& \left (\dfrac{N}{3.43\times 10^{13}\, {\rm cm}^{-2}}\right ) 
    \left( \dfrac{f}{0.4164} \right) 
    \left( \dfrac{\lambda_0}{1215.67\,\mbox{\AA}}\right) \\
    & \times \left( \dfrac{b_D}{26\,\mbox{km s}^{-1}}\right)^{-1}. \end{split}
\end{equation}
Here, $f$ is the oscillator strength, $\lambda_0$ is the rest-wavelength 
of the transition, and $b_D=\sqrt{2}v_{\rm RMS}$ is the Doppler line width, which 
we set to $26$~\kmps, the typical value measured by \citet{rudie12}.
In order to apply this to our data, we must assume that the \hone\ pixel optical depths are close to the line centre
 $\tau_{\honem} = \tau_{0,\lyam}$. This is not a bad approximation, because the steepness 
of the \hone\ column density distribution function implies that it is more likely to be close 
to the maximum optical depth of a weaker absorber than it is to be in the wing of a stronger 
absorber. Indeed, Fig.~11 of \citet{rakic12} shows that the above methodology leads to very close
agreement with the simulation result of \citet{aguirre02}. 

To convert $N_{\honem}$ to a hydrogen number density, we turn to \citet{schaye01}, who derive a relation between 
density and column density by assuming the absorbers are gravitationally 
confined gas clouds, which implies that they have sizes on the order of the local Jeans length. 
Combining their eq.~8 with eq.~\ref{eq:tau0}, we obtain:
\begin{equation}
  \begin{split}
\label{eq:nh} 
n_{\rm H} \approx& 1.19 \times 10^{-5}\,{\rm cm}^{-3} \tau_{0,\lyam}^{2/3} 
  \left( \dfrac{T}{2\times 10^4\,\mbox{K}} \right)^{0.17}\\
  & \times \left( \dfrac{\Gamma}{0.74\times10^{-12}\,\mbox{s}^{-1}} \right)^{2/3} \left( \dfrac{f_{\rm g}}{0.0154}\right)^{-1/3},
 \end{split}
\end{equation}
where we have assumed the metagalactic photoionization rate to be 
$\Gamma = 0.74\times10^{-12}\,\mbox{s}^{-1}$, the $z=2.34$ value from \citet{becker07}, and we have taken the temperature to 
be $T=2\times 10^4$~K, typical for a moderately overdense IGM region 
\cite[e.g.][]{schaye00b,lidz10,becker11,rudie12b}. The result is insensitive to the precise value of 
the temperature, as long as the gas is predominantly photoionized, as assumed by 
\citet{schaye01}. We have also assumed a gas fraction
close to the universal value of $f_{\rm g}=\Omega_{\rm b}/\Omega_{\rm m}$. 

Thus, we can use eq.~\ref{eq:nh} to convert between \hone\ optical depth
and hydrogen number density. Next, we consider the contour along a constant temperature of
$T=2\times10^4$~K in Fig.~\ref{fig:cloudy_a}, from which we obtain theoretical
optical depth ratios as a function of $n_{\text{H}}$. With eq.~\ref{eq:nh}, 
we can convert the $n_{\text{H}}$ dependence to corresponding values of $\tau_{\honem}$. 
The result of doing so is shown by the blue lines in Fig.~\ref{fig:cloudy_a}, where the different  
line styles demonstrate the effect of varying the metallicity using eq.~\ref{eq:metallicity}.

\begin{figure*}
 \includegraphics[width=\wc]{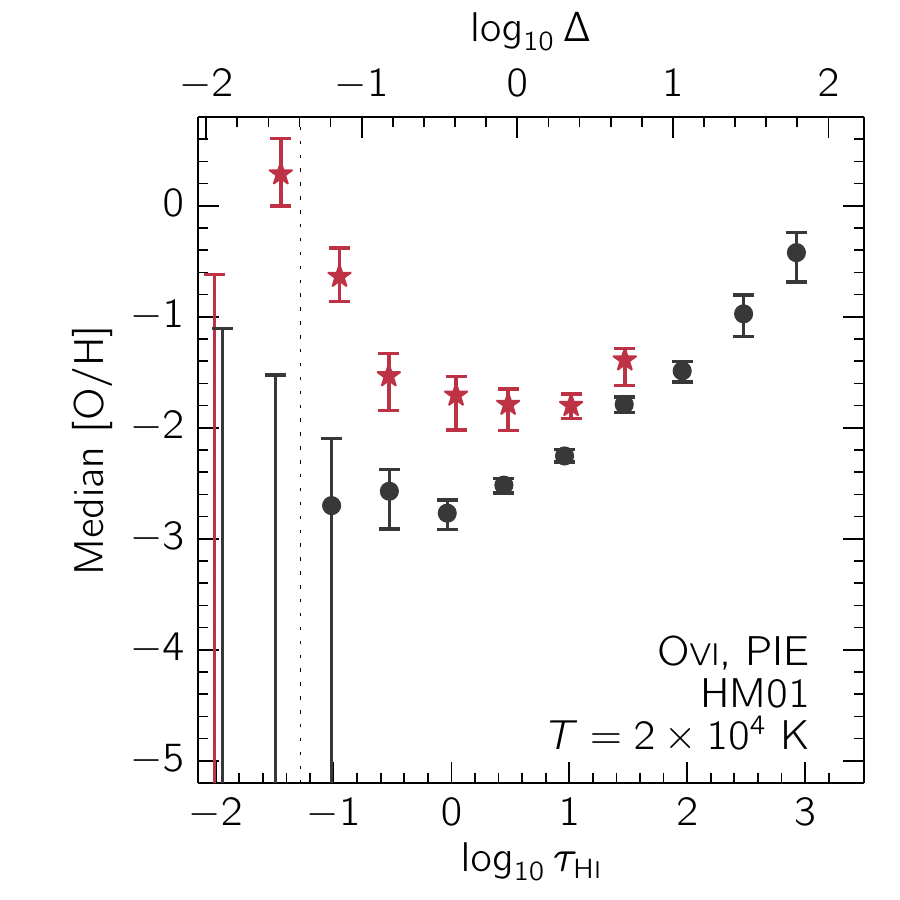}
 \includegraphics[width=\wc]{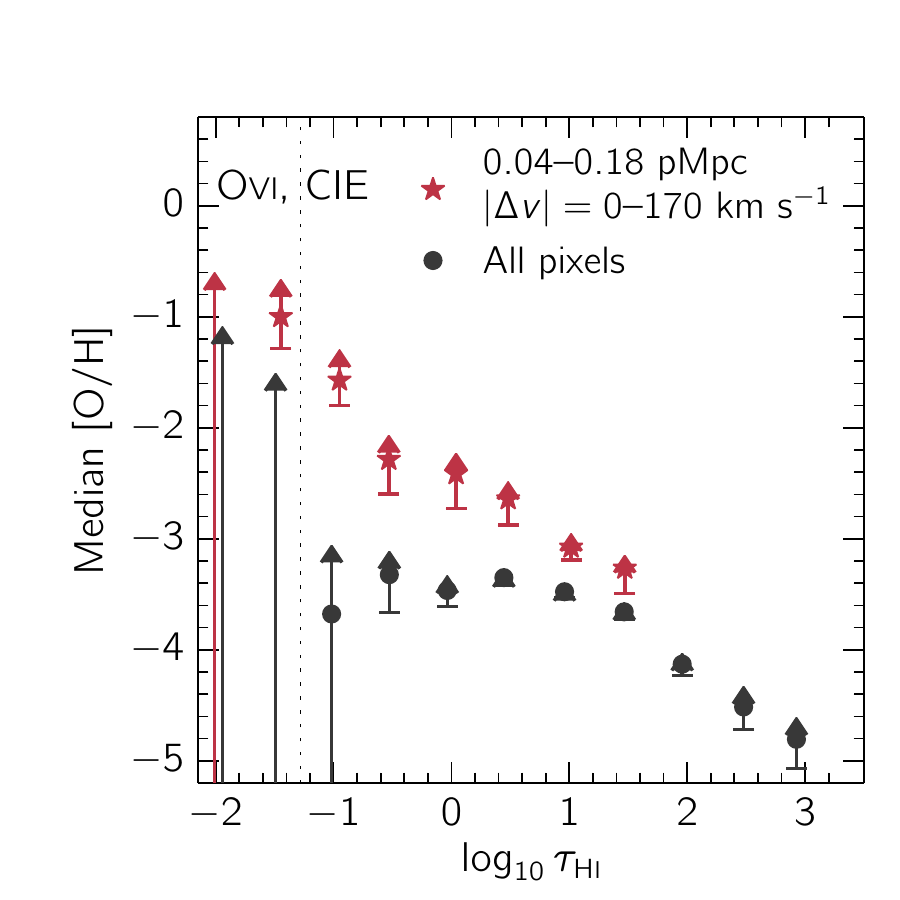} \\
 \caption{\textit{Left panel:} Metallicity as a function of \hone\ pixel optical depth. The metallicity was 
 inferred from the ratio of $\tau_{\osixm}$ to $\tau_{\honem}$ (Fig.~\ref{fig:o6_vs_h1_b})
 assuming photoionization equilibrium (PIE, $T=2\times 10^4$~K) as described in the text. On the upper axis 
  we show $\tau_{\honem}$ converted to an overdensity $\Delta$ using
  the relation for photoionized gas that we assumed in the ionization models. Random regions (black circles)
  show a positive correlation between metallicity and density, while at small galactocentric distances 
  (red stars) for $\tau_{\honem}\lesssim1$ the metallicity increases inversely with density. Such an inverted metallicity-density 
  relation is not physically expected and indicates that the assumption of PIE is probably incorrect. 
\textit{Right panel:} Lower limits on the metallicity as a function $\tau_{\honem}$, inferred again from 
the measured optical depth ratios but now assuming collisional
 ionization equilibrium (CIE). The size of each arrow is proportional to the size of the error bar on the 
 measured optical depth ratio. The inferred limits on the metallicity are lower than for PIE, but still 
 imply that at least some of the gas near galaxies is substantially enriched. }
 \label{fig:o6_vs_h1_c}
\end{figure*}

Hence, by interpolating between the blue curves in the right panel of Fig.~\ref{fig:o6_vs_h1_b},
we can estimate a metallicity (assuming PIE) for each optical depth ratio 
$\tau_{\osixm}/\tau_{\honem}$ in the left panel.
The outcome of this procedure is plotted in the left panel of Fig.~\ref{fig:o6_vs_h1_c}. To aid the 
interpretation, we indicate the baryon overdensity inferred from $\tau_{\honem}$ along the upper x-axis 
(again, we can only convert \hone\ optical depth to overdensity under the assumption of PIE). 

Focusing first on the black points that correspond to the full pixel sample, we find that
metallicity increases with overdensity $\Delta$, and is in agreement with previous pixel optical 
depth studies \citep[e.g.][]{schaye03, aguirre08}. 
Next, examining the values derived from pixels with small galactocentric 
distance (red stars), for pixels with $\tau_{\honem}\gtrsim10$ we infer
the same metallicity-density relation as for the full pixel sample, but with 
metallicities that are $\sim0.5$~dex higher. We conclude that for these \hone\ optical
depths, photoionized gas that is enriched with respect to random regions of the same density
is consistent with our 
observations of \osix(\hone) at small galactocentric distances. 

However, turning to pixels with  $\tau_{\honem}\lesssim1$, it is clear 
that the relation between metallicity and density is inverted. 
In this regime, the metallicity appears to increase 
with decreasing overdensity, with the highest metallicities 
found at the lowest overdensities, even reaching supersolar values. 
We do not consider this enrichment pattern to be physically plausible. 
Although some regions of relatively low density can be highly
enriched, they usually arise in hot superbubbles. Furthermore, an 
underdense gas phase close to galaxies with $T\sim10^4$~K
and supersolar metallicities is not consistent with 
predictions from cosmological hydrodynamical simulations 
\citep[e.g.][]{ford13, shen13}. The above suggests 
that one or more of our assumptions must be incorrect for the 
small galactocentric distance pixel sample.

\subsection{Photoionization by stars in nearby galaxies}

\begin{figure}
  \includegraphics[width=0.45\textwidth]{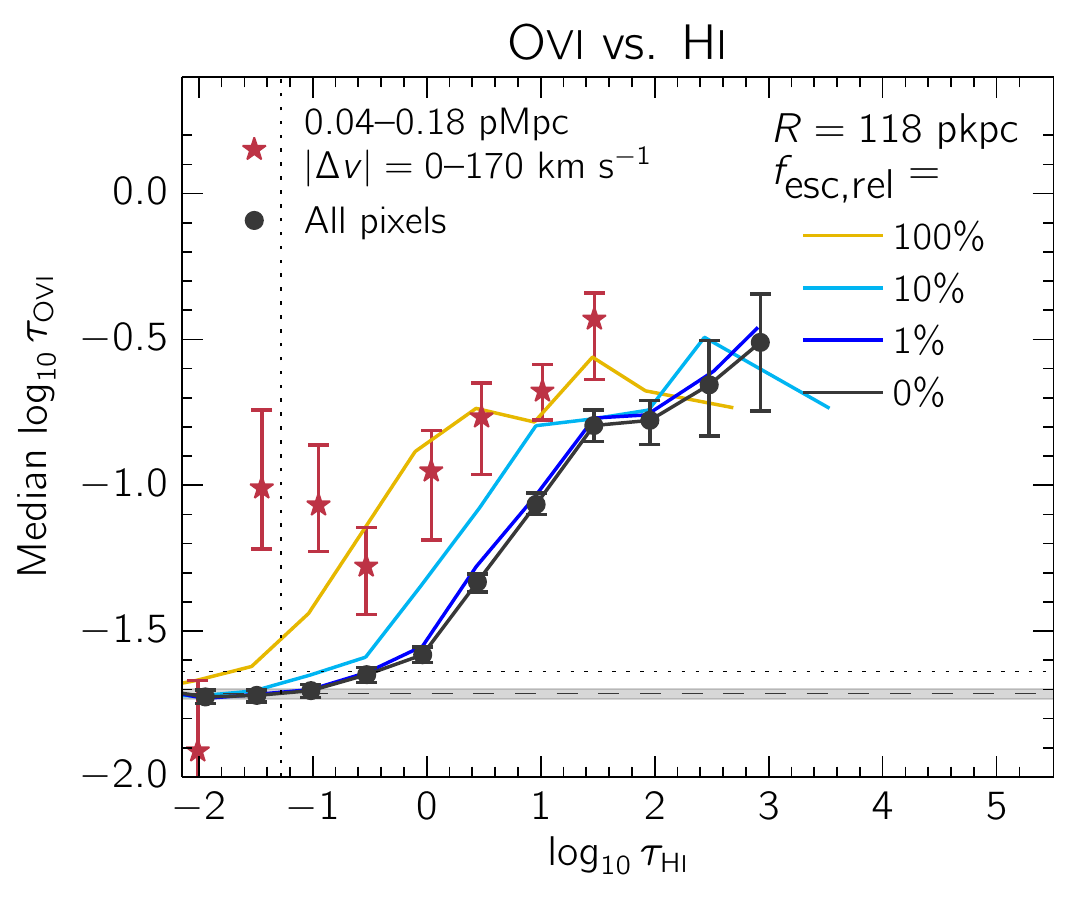} 
  \caption{The data points are identical to those in the left panel of 
  Fig.~\ref{fig:o6_vs_h1_a}. The curves show the result for the 
  full pixel sample (black circles) after dividing the \hone\ optical
depths by the boost factor in the \hone\ photoionization rate expected due to 
ionizing radiation from the nearby galaxy for the relative escape fractions,
$f_{\text{esc,rel}} \equiv f_{\text{esc},\lambda 900}/f_{\text{esc},\lambda 1500}$, 
indicated in the legend and for a distance to the source equal to the median impact 
parameter of the pixels in the small galactocentric distance sample. Even for 
$f_{\rm esc,rel}=100$\%, which is unrealistically high, the curve falls short of the 
red stars, implying that photoionization by radiation from the observed galaxies 
cannot explain the observed enhancement in \osix/\hone\ near galaxies.
}
 \label{fig:local_sources}
\end{figure}

In this section, we test whether the enhancement of $\tau_{\osixm}$ for a fixed
$\tau_{\honem}$ near galaxies can be explained by increased
photoionization by radiation from the stars in the nearby galaxies.
At sufficiently small galactocentric distances, we would expect 
the mean hydrogen ionization rate to be dominated by galactic 
radiation rather than by the metagalactic background radiation 
\citep[e.g.][]{schaye06,rahmati13}, which would reduce the 
\hone\ optical depth at fixed density. Furthermore, since stars emit very little 
radiation with photon energies $>4$~Ryd, we do not expect the \osix\ optical 
depth to be modified by local stellar sources, given that the 
ionization energy for \ofive\ is about 8.4~Ryd. Hence, if photoionization 
by nearby galaxies were important, we would expect the red stars 
in Fig.~\ref{fig:o6_vs_h1_a} to be shifted upwards relative to the 
black points and the red stars in Fig.~\ref{fig:o6_vs_h1_b} to be 
shifted towards the top-left relative to the black points. This is
qualitatively consistent with the observations, which suggests that
the difference between the two samples may be due to photoionization by
local sources rather than due to the presence of hot, enriched gas. However, 
as we will show below, quantitatively this scenario does not work out. 

We first need to evaluate the photoionization rate due to local sources, $\Gamma_{\nu_0}$,
at $\nu_0=1$~Ryd,
 \begin{equation}
  \label{eq:gamma}
  \Gamma_{\nu_0} = \int_{\nu_0}^{\infty} \dfrac{F(\nu)\sigma(\nu)}{h\nu}\, d\nu
 \end{equation}
where $h$ is the Planck constant,  $F(\nu)$ is the incident flux, and
$\sigma(\nu)$ is the photoionization cross section, taken to be
$\sigma(\nu)=\sigma_0(\nu/\nu_0)^{-3}$ where $\sigma_0 = 6.3 \times 10^{-18}$~cm$^{-2}$. 
To estimate the incident flux from a nearby galaxy, 
we assume that the galaxy is an isotropically emitting point source:
\begin{equation}
\label{eq:flux}
F(\nu) = f_{\text{esc},\nu} \dfrac{L(\nu)}{ 4 \pi R^2}
\end{equation}
where $L(\nu)$ is the galaxy luminosity, $f_{\text{esc},\nu}$ is the escape fraction of photons 
at frequency $\nu$, and $R$ is the proper distance from the point source. For the luminosity, 
we use measurements of the KBSS galaxy luminosity function from \citet{reddy09}, 
who obtained $M_{1700\text{\AA},*}=-20.70$~mag. This corresponds to
 $L_{\nu,*} = 4.05 \times 10^{28}~f_{{\rm esc},\lambda1700}^{-1}\, h_{0.7}^{-2}$~erg~s$^{-1}$~Hz$^{-1}$ at a 
 rest wavelength of $\sim1700$~\AA, where $f_{{\rm esc},\lambda1700}$ is the escape fraction at this wavelength 
 and $h_{0.7}$ is the Hubble constant in units of 70~km$\,$s$^{-1}\,$Mpc$^{-1}$. 

We then use the fact that spectral synthesis models corresponding 
to typical KBSS galaxy metallicities of 
$12+\log(\text{O/H})=8.4$ can be well-approximated by a blackbody 
curve with $T_{\text{eff}}=50000$~K \citep{steidel14}. Integrating this function numerically,
we obtain:
\begin{equation}
  \label{eq:gammaf}
  \Gamma_{\nu_0} = 14.8 \times 10^{-12} \, \text{s}^{-1} \left(\dfrac{R}{118\,\text{pkpc}} \right)^{-2}
    \dfrac{f_{\text{esc},\lambda 912}}{f_{\text{esc},\lambda 1700}}. 
\end{equation}
where we take $R=118$~pkpc because it is the median impact parameter
of the galaxies that comprise our small galactocentric distance sample. Note that this choice is conservative 
since it assumes that the absorption within $\pm 170~{\rm km}\,{\rm s}^{-1}$ of the galaxy arises in gas that 
is precisely at the distance of the galaxy. In reality, some of this gas will be in front or behind the galaxy 
and $R$ will be greater than the impact parameter.

The remaining unknown in our estimate of the photoionization rate
from local galaxies comes from the ratio of the escape 
fractions at 1~Ryd and at the observed wavelength of 1700~\AA,
$f_{\text{esc},\lambda 912}/f_{\text{esc},\lambda 1700}$.
The literature contains many values for the escape fraction of 
Lyman continuum photons relative to that of non-ionizing UV continuum photons,
$f_{\text{esc,rel}} \equiv f_{\text{esc},\lambda 900}/f_{\text{esc},\lambda 1500}$.
Values range from $\sim1$\% to $83$\% \citep[e.g.][]{steidel01, shapley06, iwata09, nestor13, mostardi13, cooke14},
although typical values are closer to the lower limit of this range.

Since we already expect  $\Gamma_{\nu_0} = 0.74 \times 10^{-12}$~s$^{-1}$ from the extragalactic
background \citep{becker07}, adding a contribution from nearby galaxies would boost the \hone\ photoionization rate by 
a factor of $\sim20$ for an escape fraction $f_{\rm esc,rel}= 100$\% at the median impact parameter and would hence suppress 
the \hone\ optical depths by the same amount. To test whether ionizing radiation from nearby galaxies can explain 
the difference between the \osix(\hone) relations of the small galactocentric radii and the random pixel samples, we have 
re-calculated the \osix(\hone) relation for random regions after  dividing all the \hone\ pixel optical depths 
by the factor by which local sources boost the \hone\ ionization rate for various values of the relative 
escape fraction. The results are compared with the observed relation for the small galactocentric radii 
sample in Fig.~\ref{fig:local_sources}. For $f_{\rm esc,rel}\le 10$\% the effect of local 
sources is not strong enough to reproduce the observations. Even for the highly 
unrealistic case that $f_{\rm esc,rel}=100$\%, the boost in the 
\hone\ ionization rate is insufficient to completely account for the 
observed enhancement of \osix\ small galactocentric distances for $\tau_{\honem}\lesssim1$.

On the other hand, while for $\tau_{\honem}\gtrsim10$ it appears that relative
escape fractions of $\sim50$\% may be able to explain our observations, 
we reiterate that \textit{average} observed relative escape fraction values 
are usually much lower than this for galaxies such as ours \citep[e.g.,][]{shapley06}. Furthermore, we note that by considering
only the transverse rather than 3-dimensional distance to the galaxy, our estimate of the
\hone\ photoionization rate boost is strictly an upper limit, and the true strength of the effect is almost
certainly smaller. 

As a final test, we have taken the blackbody spectrum used to approximate the local
galaxy radiation and used it as input, along with the extragalactic background, into
\texttt{CLOUDY}. We generated temperature-density planes showing 
optical depth ratios $\tau_{\osixm}/\tau_{\honem}$
(as in the right panel of Fig.~\ref{fig:o6_vs_h1_b}), 
as well as $\tau_{\osixm}/\tau_{\cfourm}$ and $\tau_{\osixm}/\tau_{\sifourm}$, 
and compared these with the output using only the extragalactic background.
For the \osix(\hone) relation, we indeed found that the predicted ratios
for photoionized gas at $\sim10^4$~K were higher. However, as expected
due to the fact that our assumed spectrum does not contain many photons with energies above 1~Ryd, 
for the observed ratio values there was almost no discernible difference
between the \osix(\cfour) and \osix(\sifour) relations derived
with and without the contribution from nearby galaxies.\footnote{Invoking galaxy spectra
with a soft X-ray component could potentially lead to more oxygen ionized to \osix\ \citep[e.g.,][]{cantalupo10},
however these spectral models remain uncertain and we leave testing of this scenario to a future work.}
Hence, if local sources were responsible for the enhancement in \osix(\hone), we would not 
expect to see any difference between the small galactocentric distance and random pixel samples for the 
\osix(\cfour) and \osix(\sifour) relations.
This is not consistent with the fact that we find significantly enhanced
\osix\ for both fixed \cfour\ and fixed \sifour\ much like we do for the \osix(\hone) relation
(see the centre and right panels of Fig.~\ref{fig:o6_vs_h1_a}).
 Thus, we conclude that enhanced photoionization due to the proximity of 
 the observed galaxies cannot explain the enhancement 
of \osix\ at fixed \hone\ that we observe for
 the small galactocentric distance pixel sample.

 In the above scenario, we only consider photoionization by stars in the nearby galaxies because there is no evidence that our 
galaxies contain AGN. However, it was recently pointed out by \citet{oppenheimer13b} that because of the long recombination 
times for metals at densities typical of the circumgalactic medium, ions like \osix\ may remain out of 
ionization equilibrium long after the AGN episode is over. In fact, for reasonable AGN duty cycles, the authors argue
that much of the \osix\ detected in quasar spectra resides in such fossil AGN proximity zones. On the other hand, in regions 
where the equilibrium neutral hydrogen fraction is low, \hone\ will equilibrate nearly instantaneously after the AGN turns off.
Hence, in this scenario \hone\ would be photoionized and in equilibrium, while \osix\ is out of equilibrium and enhanced 
because of the fossil AGN proximity effect. We would then expect the enhancement in \osix\ to be strongest in regions of 
low \hone\ absorption, since such regions correspond to low densities and thus long recombination times. 
This is in qualitative agreement with our observations, although \citet{oppenheimer13b} show that at very low
density oxygen will become so highly ionized that \osix\ is suppressed, which would be inconsistent
with our findings. 

More generally, since (fossil) AGN proximity and higher gas temperatures both tend to increase the abundance of more highly 
ionized species relative to ions with lower ionization potentials, many of the predictions of these two scenarios will be 
qualitatively similar, although it remains to be seen whether the fossil proximity effect can work quantitatively. 
One important difference, however, is the expected widths of the absorption lines. 
If the oxygen is hot enough for \osix\ to be collisionally ionized, then the absorption lines will be broader than 
if the gas were photoionized. We intend to measure the widths of the metal absorption lines and to model the fossil AGN 
proximity effect in future work.

\subsection{Collisionally ionized gas}
\label{sec:cie}

It may be that at small galactocentric distances, 
much of the \osix\ absorption arises in collisionally ionized 
rather than photoionized gas. Indeed, the behaviour of the \osix(\hone) relation
for the near-galaxy pixels with $\tau_{\honem}\lesssim1$ points to collisional ionization playing a role. 
In general, we would not expect to see very much \hone\ associated with 
collisionally ionized \osix, as the \hone\ fraction is 
very low at $T> 10^5$~K. The behaviour of the [O/H]($\Delta$) 
relation that we inferred under the assumption of photoionization (Fig.~\ref{fig:o6_vs_h1_c}) 
is rather suggestive, as the relation displays larger 
departures from the random regions with decreasing $\tau_{\honem}$.

If we are in fact probing collisionally ionized gas, then we cannot measure its 
metallicity as was done in \S~\ref{sec:pie} -- first because we are unable to estimate the gas density from 
the \hone\ optical depth, and second because there is no natural equilibrium temperature.
However, it is possible to put a lower
limit on the metallicity by selecting the temperature and density that would maximize 
$\tau_{\osixm}/\tau_{\honem}$. In Fig.~\ref{fig:cloudy_a}, we 
have marked the temperature and density where this theoretical maximum is reached 
by a small black circle.
Then, in the right-hand panel of Fig.~\ref{fig:o6_vs_h1_b}, the horizontal orange 
lines show the $\tau_{\osixm}/\tau_{\honem}$ ratio at this point for different metallicities.

Thus, irrespective of \hone\ optical depth (which has no correspondence with
density under the assumption of CIE) we can convert the observed optical
depth ratios in the left-hand panel of Fig.~\ref{fig:o6_vs_h1_b} to 
lower-limits on the metallicity, and we show the result in the right panel 
of Fig.~\ref{fig:o6_vs_h1_c}. As expected, the limits obtained 
are lower than the metallicities inferred when assuming photoionization. 
Nevertheless, some of the lower limits are sufficiently high to be interesting. 
The $1\sigma$ lower limit approaches $\text{[O/H]}\sim-1$ at the lowest 
bins in \hone\ optical depth, where $\log_{10}\tau_{\honem}=-1.5$. 

Given the reasonable metallicity limits 
in Fig.~\ref{fig:o6_vs_h1_c} combined
with the fact that the previous two scenarios (photoionization by 
the extragalactic background or by local stellar radiation) cannot
account for the \osix\ enhancement for $\tau_{\honem}\lesssim1$, 
we conclude that collisional ionization is the dominant ionization 
source for a significant fraction of the small galactocentric distance pixels. 
We emphasize that in collisionally ionized gas, $\tau_{\honem}$ is 
no longer a good estimator of density, and we could potentially 
be probing very high overdensities.

We also examined the \osix/\cfour\ and \osix/\sifour\ ratios 
in Appendix~\ref{sec:o6_vs_c4_and_si4}.
We conclude that \osix\ pixels near galaxies are more
consistent with a hotter gas phase than the full pixel sample, and furthermore we 
find that the strength of this trend increases inversely with \cfour\ and \sifour\ optical depths,
as might be expected for collisionally ionized gas at $\sim 3\times10^5$~K where these other ions are not abundant. 
However, since the results are not constraining, the details are left to the Appendix.

\section{Discussion and conclusions}
\label{sec:conclusion}

We have used absorption by \hone\ and metals to study the physical conditions near $z\sim 2.3$ 
star-forming galaxies in the fields of 15 hyper-luminous background QSOs that have been observed with 
Keck/HIRES as part of the Keck Baryonic Structure Survey (KBSS). We focused on \ngaltotalSS\ galaxies
with impact parameters $<180$~pkpc 
and isolated the pixels of the quasar spectra that are within $\pm170$~\kmps\ of the accurate galaxy redshifts provided by the KBSS. 
In \citet{turner14} we showed that the absorption by \hone\ and various metal ions is strongly enhanced in these circumgalactic 
regions. The fact that both \hone\ and the metals are enhanced raises the question whether the observed increase in the metal 
absorption merely reflects the presence of higher gas densities near galaxies or whether it implies that the gas near galaxies 
has a higher metallicity at fixed density or a different temperature from gas in random regions.  

To address this question, we measured the pixel optical depths of \osix\ as a function of \hone, \cfour\ and \sifour, 
and compared the results for the pixels located at small galactocentric distances to 
the full pixel sample, which is representative of random regions. 
Our main result is the detection of a 0.3--0.7~dex enhancement (which reaches its maximum at $\log_{10} \tau_{\honem}\sim-1.5$)
in the median optical depth of \osix\ at fixed $\tau_{\honem}$ for the small galactocentric
distance sample when compared with the full pixel sample (Fig.~\ref{fig:o6_vs_h1_a}). We verified that 
this enhancement, which we detected at 2--3$\sigma$ confidence per logarithmic bin for 
$\tau_{\honem}$, $\tau_{\cfourm}$, and $\tau_{\sifourm}$, is not due to differences 
in the redshift distribution or the quality of the quasar spectra between the small galactocentric distance and 
full pixel samples. 

We proposed and tested three different hypotheses that may explain the observed enhancement of \osix(\hone) near galaxies:
(1) the gas is photoionized by the extragalactic background but has a higher metallicity at fixed density; 
(2) the gas is more highly ionized at fixed density by radiation from stars in the nearby galaxy;
and (3) the enriched gas is hot and collisionally ionized.

To test scenario (1), we used \texttt{CLOUDY} ionization models and the relation between $\tau_{\honem}$ and 
density for photoionized, self-gravitating clouds from \citet{schaye01} and \citet{rakic12} to convert the 
observed optical depth ratios into a metallicity-density relation, assuming $T\sim 10^4\,$K, as expected for photoionized gas. 
We found that the full pixel sample gave a physically plausible metallicity-density relation that is consistent 
with previous studies which did not have information on the locations of galaxies. 
Furthermore, the same metallicity-density relation (but shifted up by $\sim0.5$~dex in metallicity) is also present 
for small galactocentric distance pixels that have $\tau_{\honem}\gtrsim10$.
Thus, for high \hone\ optical depths, the enhancement of \osix\ at fixed $\tau_{\honem}$
is consistent with arising predominantly from enriched, photoionized gas.
However, making this same assumption of PIE for $\tau_{\honem}\lesssim1$
resulted in an [O/H] versus overdensity relation that cannot be easily explained,
as [O/H] was found to increase strongly with decreasing overdensity, implying supersolar metallicities for 
underdense gas (left panel of Fig.~\ref{fig:o6_vs_h1_c}). We therefore concluded that while photoionization by the 
background radiation is a plausible scenario for the random regions, it cannot account for the 
observed enhancement of \osix(\hone) near galaxies.

In scenario (2) \osix\ is enhanced at fixed \hone\ because radiation from stars in the nearby galaxy suppresses \hone, 
while leaving \osix\ unchanged.  To test this explanation, we estimated the \hone\ photoionization rate due to 
the galaxies. However, we found that only under the unrealistic assumptions that the relative escape fraction 
$f_{\text{esc},\lambda 900}/f_{\text{esc},\lambda 1500}$ is 100\% and that the 3-D distance between the gas and 
the galaxy is equal to the impact parameter, can the flux of ionizing radiation from the galaxies explain 
the observed increase in $\tau_{\osixm}/\tau_{\honem}$ near galaxies. Reducing the relative
escape fraction to a still conservative value of 10\% rules out this scenario as a dominant ionization source
(Fig.~\ref{fig:local_sources}). Furthermore, we found that such 
a galaxy proximity effect is predicted to have a much smaller effect on \osix(\cfour) and \osix(\sifour) than on \osix(\hone), 
which is inconsistent with our observations.

Contrary to photoionization by either the extragalactic background or local stellar radiation, scenario (3) can explain the 
enhancement in \osix(\hone) near galaxies for pixels with $\tau_{\honem}\lesssim1$. 
If a substantial fraction of the enriched gas near galaxies is sufficiently 
hot for \osix\ to be collisionally ionized, i.e.\ $T> 10^5\,$K, then we can account for the observations. By assuming the 
maximum \osix/\hone\ ratio reached in collisional ionization equilibrium, we converted the observed optical depth ratios into 
lower limits on the metallicity, finding ${\rm [O/H]} \gtrsim -1$ for gas with weak \hone\ absorption 
(right panel of Fig.~\ref{fig:o6_vs_h1_c}). 
Indeed, this measurement is supported by other characterizations of KBSS galaxy properties. \citet{rudie12} found 
higher temperatures for fixed $\tau_{\honem}$ near the KBSS galaxies, while \citet{steidel14} measured \htwo\ region 
metallicities of $0.4$~Z$_{\odot}$, which could serve as a possible upper limit to the value presented	 here
(although if metal-enriched winds drive most of the metals out of 
the star-forming regions, it is possible 
that the circumgalactic gas may have a higher metallicity than galactic \htwo\ regions). 
Furthermore, the inferred metallicities  and temperatures of $T\gtrsim10^5$~K are in agreement with the 
predictions of \citet{vandevoort12} for galaxies with the masses 
and redshifts similar to ours.

In summary, we favour the conclusion
that our galaxies are surrounded by hot ($T > 10^5\,$K) gas of which a substantial fraction must have metallicity 
$\gtrsim 10^{-1}$ of solar. Furthermore, we find that this metal-enriched phase extends 
out to $\sim350$~\kmps\ of the galaxy 
positions (Fig.~\ref{fig:vary_vel}), which corresponds to $\gtrsim1.5$ times the halo circular velocities. 
Because of the relatively high temperature that requires shock-heating, 
the large velocity range extending far outside the haloes, and high metallicity, we conclude that we have detected hot, 
metal-enriched outflowing gas. 
Future comparisons with hydrodynamical simulations, considering ion ratios as well as the kinematics and line widths, 
will provide strong constraints on models of galaxy formation and may provide further insight into the interpretation of our observations. 

\section*{Acknowledgements}

We are very grateful to Milan Bogosavljevic, Alice Shapley, Dawn Erb, Naveen Reddy, Max Pettini, Ryan Trainor, 
and David Law for their invaluable contributions to the Keck Baryonic Structure Survey, without which the results 
presented here would have not been possible. We also thank Ryan Cooke for his help with the continuum fitting of 
QSO spectra. We gratefully acknowledge support from Marie Curie Training Network CosmoComp (PITN-GA-2009- 238356) 
and the European Research Council under the European Union's Seventh Framework Programme (FP7/2007-2013)/ERC Grant 
agreement 278594-GasAroundGalaxies. CCS, GCR, ALS acknowledge support from grants AST-0908805 and AST-13131472 
from the US National Science Foundation.
This work is based on data obtained
at the W.M. Keck Observatory, which is operated as 
a scientific partnership among the California Institute of 
Technology, the University of California, and NASA, and was made possible
by the generous financial support of the W.M. Keck Foundation.
We thank the W. M. Keck Observatory staff for their assistance with the 
observations. We also thank the Hawaiian people, as without their hospitality the observations presented here 
would have not been possible.


\bibliographystyle{apj} 
\bibliography{bibliography}

\begin{thebibliography}{}
\expandafter\ifx\csname natexlab\endcsname\relax\def\natexlab#1{#1}\fi

\bibitem[{{Adelberger} {et~al.}(2005{\natexlab{a}}){Adelberger}, {Shapley},
  {Steidel}, {Pettini}, {Erb}, \& {Reddy}}]{adelberger05b}
{Adelberger}, K.~L., {Shapley}, A.~E., {Steidel}, C.~C., {et~al.}
  2005{\natexlab{a}}, \apj, 629, 636

\bibitem[{{Adelberger} {et~al.}(2005{\natexlab{b}}){Adelberger}, {Steidel},
  {Pettini}, {Shapley}, {Reddy}, \& {Erb}}]{adelberger05a}
{Adelberger}, K.~L., {Steidel}, C.~C., {Pettini}, M., {et~al.}
  2005{\natexlab{b}}, \apj, 619, 697

\bibitem[{{Adelberger} {et~al.}(2004){Adelberger}, {Steidel}, {Shapley},
  {Hunt}, {Erb}, {Reddy}, \& {Pettini}}]{adelberger04}
{Adelberger}, K.~L., {Steidel}, C.~C., {Shapley}, A.~E., {et~al.} 2004, \apj,
  607, 226

\bibitem[{{Adelberger} {et~al.}(2003){Adelberger}, {Steidel}, {Shapley}, \&
  {Pettini}}]{adelberger03}
{Adelberger}, K.~L., {Steidel}, C.~C., {Shapley}, A.~E., \& {Pettini}, M. 2003,
  \apj, 584, 45

\bibitem[{{Aguirre} {et~al.}(2008){Aguirre}, {Dow-Hygelund}, {Schaye}, \&
  {Theuns}}]{aguirre08}
{Aguirre}, A., {Dow-Hygelund}, C., {Schaye}, J., \& {Theuns}, T. 2008, \apj,
  689, 851

\bibitem[{{Aguirre} {et~al.}(2002){Aguirre}, {Schaye}, \& {Theuns}}]{aguirre02}
{Aguirre}, A., {Schaye}, J., \& {Theuns}, T. 2002, \apj, 576, 1

\bibitem[{{Allende Prieto} {et~al.}(2001){Allende Prieto}, {Lambert}, \&
  {Asplund}}]{allende01}
{Allende Prieto}, C., {Lambert}, D.~L., \& {Asplund}, M. 2001, \apjl, 556, L63

\bibitem[{{Allende Prieto} {et~al.}(2002){Allende Prieto}, {Lambert}, \&
  {Asplund}}]{allende02}
---. 2002, \apjl, 573, L137

\bibitem[{{Aracil} {et~al.}(2004){Aracil}, {Petitjean}, {Pichon}, \&
  {Bergeron}}]{aracil04}
{Aracil}, B., {Petitjean}, P., {Pichon}, C., \& {Bergeron}, J. 2004, \aap, 419,
  811

\bibitem[{{Becker} \& {Bolton}(2013)}]{becker13}
{Becker}, G.~D., \& {Bolton}, J.~S. 2013, \mnras, 436, 1023

\bibitem[{{Becker} {et~al.}(2011){Becker}, {Bolton}, {Haehnelt}, \&
  {Sargent}}]{becker11}
{Becker}, G.~D., {Bolton}, J.~S., {Haehnelt}, M.~G., \& {Sargent}, W.~L.~W.
  2011, \mnras, 410, 1096

\bibitem[{{Becker} {et~al.}(2007){Becker}, {Rauch}, \& {Sargent}}]{becker07}
{Becker}, G.~D., {Rauch}, M., \& {Sargent}, W.~L.~W. 2007, \apj, 662, 72

\bibitem[{{Bergeron} {et~al.}(2002){Bergeron}, {Aracil}, {Petitjean}, \&
  {Pichon}}]{bergeron02}
{Bergeron}, J., {Aracil}, B., {Petitjean}, P., \& {Pichon}, C. 2002, \aap, 396,
  L11

\bibitem[{{Cantalupo}(2010)}]{cantalupo10}
{Cantalupo}, S. 2010, \mnras, 403, L16

\bibitem[{{Carswell} {et~al.}(2002){Carswell}, {Schaye}, \& {Kim}}]{carswell02}
{Carswell}, B., {Schaye}, J., \& {Kim}, T.-S. 2002, \apj, 578, 43

\bibitem[{{Chen} \& {Mulchaey}(2009)}]{chen09}
{Chen}, H.-W., \& {Mulchaey}, J.~S. 2009, \apj, 701, 1219

\bibitem[{{Conroy} {et~al.}(2008){Conroy}, {Shapley}, {Tinker}, {Santos}, \&
  {Lemson}}]{conroy08}
{Conroy}, C., {Shapley}, A.~E., {Tinker}, J.~L., {Santos}, M.~R., \& {Lemson},
  G. 2008, \apj, 679, 1192

\bibitem[{{Cooke} {et~al.}(2014){Cooke}, {Ryan-Weber}, {Garel}, \&
  {D{\'{\i}}az}}]{cooke14}
{Cooke}, J., {Ryan-Weber}, E.~V., {Garel}, T., \& {D{\'{\i}}az}, C.~G. 2014,
  \mnras, 441, 837

\bibitem[{{Cowie} \& {Songaila}(1998)}]{cowie98}
{Cowie}, L.~L., \& {Songaila}, A. 1998, \nat, 394, 44

\bibitem[{{Danforth} \& {Shull}(2008)}]{danforth08}
{Danforth}, C.~W., \& {Shull}, J.~M. 2008, \apj, 679, 194

\bibitem[{{Dekel} {et~al.}(2009){Dekel}, {Birnboim}, {Engel}, {Freundlich},
  {Goerdt}, {Mumcuoglu}, {Neistein}, {Pichon}, {Teyssier}, \&
  {Zinger}}]{dekel09}
{Dekel}, A., {Birnboim}, Y., {Engel}, G., {et~al.} 2009, \nat, 457, 451

\bibitem[{{Ellison} {et~al.}(2000){Ellison}, {Songaila}, {Schaye}, \&
  {Pettini}}]{ellison00}
{Ellison}, S.~L., {Songaila}, A., {Schaye}, J., \& {Pettini}, M. 2000, \aj,
  120, 1175

\bibitem[{{Erb} {et~al.}(2006{\natexlab{a}}){Erb}, {Steidel}, {Shapley},
  {Pettini}, {Reddy}, \& {Adelberger}}]{erb06b}
{Erb}, D.~K., {Steidel}, C.~C., {Shapley}, A.~E., {et~al.} 2006{\natexlab{a}},
  \apj, 647, 128

\bibitem[{{Erb} {et~al.}(2006{\natexlab{b}}){Erb}, {Steidel}, {Shapley},
  {Pettini}, {Reddy}, \& {Adelberger}}]{erb06c}
---. 2006{\natexlab{b}}, \apj, 646, 107

\bibitem[{{Faucher-Gigu{\`e}re} {et~al.}(2008){Faucher-Gigu{\`e}re}, {Lidz},
  {Hernquist}, \& {Zaldarriaga}}]{fauchergiguere08}
{Faucher-Gigu{\`e}re}, C.-A., {Lidz}, A., {Hernquist}, L., \& {Zaldarriaga}, M.
  2008, \apj, 688, 85

\bibitem[{{Ferland} {et~al.}(2013){Ferland}, {Porter}, {van Hoof}, {Williams},
  {Abel}, {Lykins}, {Shaw}, {Henney}, \& {Stancil}}]{ferland13}
{Ferland}, G.~J., {Porter}, R.~L., {van Hoof}, P.~A.~M., {et~al.} 2013, \rmxaa,
  49, 137

\bibitem[{{Ford} {et~al.}(2013){Ford}, {Oppenheimer}, {Dav{\'e}}, {Katz},
  {Kollmeier}, \& {Weinberg}}]{ford13}
{Ford}, A.~B., {Oppenheimer}, B.~D., {Dav{\'e}}, R., {et~al.} 2013, \mnras,
  432, 89

\bibitem[{{Fox} {et~al.}(2008){Fox}, {Bergeron}, \& {Petitjean}}]{fox08}
{Fox}, A.~J., {Bergeron}, J., \& {Petitjean}, P. 2008, \mnras, 388, 1557

\bibitem[{{Haardt} \& {Madau}(2001)}]{haardt01}
{Haardt}, F., \& {Madau}, P. 2001, in Clusters of Galaxies and the High
  Redshift Universe Observed in X-rays, ed. D.~M. {Neumann} \& J.~T.~V. {Tran}

\bibitem[{{Haardt} \& {Madau}(2012)}]{haardt12}
{Haardt}, F., \& {Madau}, P. 2012, \apj, 746, 125

\bibitem[{{Holweger}(2001)}]{holweger01}
{Holweger}, H. 2001, in American Institute of Physics Conference Series, Vol.
  598, Joint SOHO/ACE workshop ''Solar and Galactic Composition'', ed. R.~F.
  {Wimmer-Schweingruber}, 23--30

\bibitem[{{Iwata} {et~al.}(2009){Iwata}, {Inoue}, {Matsuda}, {Furusawa},
  {Hayashino}, {Kousai}, {Akiyama}, {Yamada}, {Burgarella}, \&
  {Deharveng}}]{iwata09}
{Iwata}, I., {Inoue}, A.~K., {Matsuda}, Y., {et~al.} 2009, \apj, 692, 1287

\bibitem[{{Kere{\v s}} {et~al.}(2005){Kere{\v s}}, {Katz}, {Weinberg}, \&
  {Dav{\'e}}}]{keres05}
{Kere{\v s}}, D., {Katz}, N., {Weinberg}, D.~H., \& {Dav{\'e}}, R. 2005,
  \mnras, 363, 2

\bibitem[{{Lidz} {et~al.}(2010){Lidz}, {Faucher-Gigu{\`e}re}, {Dall'Aglio},
  {McQuinn}, {Fechner}, {Zaldarriaga}, {Hernquist}, \& {Dutta}}]{lidz10}
{Lidz}, A., {Faucher-Gigu{\`e}re}, C.-A., {Dall'Aglio}, A., {et~al.} 2010,
  \apj, 718, 199

\bibitem[{{Lopez} {et~al.}(2007){Lopez}, {Ellison}, {D'Odorico}, \&
  {Kim}}]{lopez07}
{Lopez}, S., {Ellison}, S., {D'Odorico}, S., \& {Kim}, T.-S. 2007, \aap, 469,
  61

\bibitem[{{Mostardi} {et~al.}(2013){Mostardi}, {Shapley}, {Nestor}, {Steidel},
  {Reddy}, \& {Trainor}}]{mostardi13}
{Mostardi}, R.~E., {Shapley}, A.~E., {Nestor}, D.~B., {et~al.} 2013, \apj, 779,
  65

\bibitem[{{Nestor} {et~al.}(2013){Nestor}, {Shapley}, {Kornei}, {Steidel}, \&
  {Siana}}]{nestor13}
{Nestor}, D.~B., {Shapley}, A.~E., {Kornei}, K.~A., {Steidel}, C.~C., \&
  {Siana}, B. 2013, \apj, 765, 47

\bibitem[{{Oppenheimer} {et~al.}(2012){Oppenheimer}, {Dav{\'e}}, {Katz},
  {Kollmeier}, \& {Weinberg}}]{oppenheimer12}
{Oppenheimer}, B.~D., {Dav{\'e}}, R., {Katz}, N., {Kollmeier}, J.~A., \&
  {Weinberg}, D.~H. 2012, \mnras, 420, 829

\bibitem[{{Oppenheimer} \& {Schaye}(2013{\natexlab{a}})}]{oppenheimer13b}
{Oppenheimer}, B.~D., \& {Schaye}, J. 2013{\natexlab{a}}, \mnras, 434, 1063

\bibitem[{{Oppenheimer} \& {Schaye}(2013{\natexlab{b}})}]{oppenheimer13a}
---. 2013{\natexlab{b}}, \mnras, 434, 1043

\bibitem[{{Pieri} {et~al.}(2010){Pieri}, {Frank}, {Mathur}, {Weinberg}, {York},
  \& {Oppenheimer}}]{pieri10}
{Pieri}, M.~M., {Frank}, S., {Mathur}, S., {et~al.} 2010, \apj, 716, 1084

\bibitem[{{Pieri} {et~al.}(2006){Pieri}, {Schaye}, \& {Aguirre}}]{pieri06}
{Pieri}, M.~M., {Schaye}, J., \& {Aguirre}, A. 2006, \apj, 638, 45

\bibitem[{{Pieri} {et~al.}(2014){Pieri}, {Mortonson}, {Frank}, {Crighton},
  {Weinberg}, {Lee}, {Noterdaeme}, {Bailey}, {Busca}, {Ge}, {Kirkby},
  {Lundgren}, {Mathur}, {P{\^a}ris}, {Palanque-Delabrouille}, {Petitjean},
  {Rich}, {Ross}, {Schneider}, \& {York}}]{pieri14}
{Pieri}, M.~M., {Mortonson}, M.~J., {Frank}, S., {et~al.} 2014, \mnras, 441,
  1718

\bibitem[{{Planck Collaboration} {et~al.}(2013){Planck Collaboration}, {Ade},
  {Aghanim}, {Armitage-Caplan}, {Arnaud}, {Ashdown}, {Atrio-Barandela},
  {Aumont}, {Baccigalupi}, {Banday}, \& et~al.}]{planck13}
{Planck Collaboration}, {Ade}, P.~A.~R., {Aghanim}, N., {et~al.} 2013, ArXiv
  e-prints, arXiv:1303.5076

\bibitem[{{Prochaska} {et~al.}(2011){Prochaska}, {Weiner}, {Chen}, {Mulchaey},
  \& {Cooksey}}]{prochaska11}
{Prochaska}, J.~X., {Weiner}, B., {Chen}, H.-W., {Mulchaey}, J., \& {Cooksey},
  K. 2011, \apj, 740, 91

\bibitem[{{Rahmati} \& {Schaye}(2014)}]{rahmati14}
{Rahmati}, A., \& {Schaye}, J. 2014, \mnras, 438, 529

\bibitem[{{Rahmati} {et~al.}(2013){Rahmati}, {Schaye}, {Pawlik}, \&
  {Raicevic}}]{rahmati13}
{Rahmati}, A., {Schaye}, J., {Pawlik}, A.~H., \& {Raicevic}, M. 2013, \mnras,
  431, 2261

\bibitem[{{Rakic} {et~al.}(2013){Rakic}, {Schaye}, {Steidel}, {Booth}, {Dalla
  Vecchia}, \& {Rudie}}]{rakic13}
{Rakic}, O., {Schaye}, J., {Steidel}, C.~C., {et~al.} 2013, \mnras, 433, 3103

\bibitem[{{Rakic} {et~al.}(2011){Rakic}, {Schaye}, {Steidel}, \&
  {Rudie}}]{rakic11}
{Rakic}, O., {Schaye}, J., {Steidel}, C.~C., \& {Rudie}, G.~C. 2011, \mnras,
  414, 3265

\bibitem[{{Rakic} {et~al.}(2012){Rakic}, {Schaye}, {Steidel}, \&
  {Rudie}}]{rakic12}
---. 2012, \apj, 751, 94

\bibitem[{{Reddy} \& {Steidel}(2009)}]{reddy09}
{Reddy}, N.~A., \& {Steidel}, C.~C. 2009, \apj, 692, 778

\bibitem[{{Reddy} {et~al.}(2008){Reddy}, {Steidel}, {Pettini}, {Adelberger},
  {Shapley}, {Erb}, \& {Dickinson}}]{reddy08}
{Reddy}, N.~A., {Steidel}, C.~C., {Pettini}, M., {et~al.} 2008, \apjs, 175, 48

\bibitem[{{Rudie} {et~al.}(2012{\natexlab{a}}){Rudie}, {Steidel}, \&
  {Pettini}}]{rudie12b}
{Rudie}, G.~C., {Steidel}, C.~C., \& {Pettini}, M. 2012{\natexlab{a}}, \apjl,
  757, L30

\bibitem[{{Rudie} {et~al.}(2012{\natexlab{b}}){Rudie}, {Steidel}, {Trainor},
  {Rakic}, {Bogosavljevi{\'c}}, {Pettini}, {Reddy}, {Shapley}, {Erb}, \&
  {Law}}]{rudie12}
{Rudie}, G.~C., {Steidel}, C.~C., {Trainor}, R.~F., {et~al.}
  2012{\natexlab{b}}, \apj, 750, 67

\bibitem[{{Savage} {et~al.}(2014){Savage}, {Kim}, {Wakker}, {Keeney}, {Shull},
  {Stocke}, \& {Green}}]{savage14}
{Savage}, B.~D., {Kim}, T.-S., {Wakker}, B.~P., {et~al.} 2014, \apjs, 212, 8

\bibitem[{{Schaye}(2001)}]{schaye01}
{Schaye}, J. 2001, \apj, 559, 507

\bibitem[{{Schaye}(2006)}]{schaye06}
---. 2006, \apj, 643, 59

\bibitem[{{Schaye} {et~al.}(2003){Schaye}, {Aguirre}, {Kim}, {Theuns}, {Rauch},
  \& {Sargent}}]{schaye03}
{Schaye}, J., {Aguirre}, A., {Kim}, T.-S., {et~al.} 2003, \apj, 596, 768

\bibitem[{{Schaye} {et~al.}(2007){Schaye}, {Carswell}, \& {Kim}}]{schaye07}
{Schaye}, J., {Carswell}, R.~F., \& {Kim}, T.-S. 2007, \mnras, 379, 1169

\bibitem[{{Schaye} {et~al.}(2000{\natexlab{a}}){Schaye}, {Rauch}, {Sargent}, \&
  {Kim}}]{schaye00a}
{Schaye}, J., {Rauch}, M., {Sargent}, W.~L.~W., \& {Kim}, T.-S.
  2000{\natexlab{a}}, \apjl, 541, L1

\bibitem[{{Schaye} {et~al.}(2000{\natexlab{b}}){Schaye}, {Theuns}, {Rauch},
  {Efstathiou}, \& {Sargent}}]{schaye00b}
{Schaye}, J., {Theuns}, T., {Rauch}, M., {Efstathiou}, G., \& {Sargent},
  W.~L.~W. 2000{\natexlab{b}}, \mnras, 318, 817

\bibitem[{{Shapley} {et~al.}(2003){Shapley}, {Steidel}, {Pettini}, \&
  {Adelberger}}]{shapley03}
{Shapley}, A.~E., {Steidel}, C.~C., {Pettini}, M., \& {Adelberger}, K.~L. 2003,
  \apj, 588, 65

\bibitem[{{Shapley} {et~al.}(2006){Shapley}, {Steidel}, {Pettini},
  {Adelberger}, \& {Erb}}]{shapley06}
{Shapley}, A.~E., {Steidel}, C.~C., {Pettini}, M., {Adelberger}, K.~L., \&
  {Erb}, D.~K. 2006, \apj, 651, 688

\bibitem[{{Shen} {et~al.}(2013){Shen}, {Madau}, {Guedes}, {Mayer}, {Prochaska},
  \& {Wadsley}}]{shen13}
{Shen}, S., {Madau}, P., {Guedes}, J., {et~al.} 2013, \apj, 765, 89

\bibitem[{{Simcoe} {et~al.}(2002){Simcoe}, {Sargent}, \& {Rauch}}]{simcoe02}
{Simcoe}, R.~A., {Sargent}, W.~L.~W., \& {Rauch}, M. 2002, \apj, 578, 737

\bibitem[{{Simcoe} {et~al.}(2004){Simcoe}, {Sargent}, \& {Rauch}}]{simcoe04}
---. 2004, \apj, 606, 92

\bibitem[{{Simcoe} {et~al.}(2006){Simcoe}, {Sargent}, {Rauch}, \&
  {Becker}}]{simcoe06}
{Simcoe}, R.~A., {Sargent}, W.~L.~W., {Rauch}, M., \& {Becker}, G. 2006, \apj,
  637, 648

\bibitem[{{Steidel} {et~al.}(2003){Steidel}, {Adelberger}, {Shapley},
  {Pettini}, {Dickinson}, \& {Giavalisco}}]{steidel03}
{Steidel}, C.~C., {Adelberger}, K.~L., {Shapley}, A.~E., {et~al.} 2003, \apj,
  592, 728

\bibitem[{{Steidel} {et~al.}(2010){Steidel}, {Erb}, {Shapley}, {Pettini},
  {Reddy}, {Bogosavljevi{\'c}}, {Rudie}, \& {Rakic}}]{steidel10}
{Steidel}, C.~C., {Erb}, D.~K., {Shapley}, A.~E., {et~al.} 2010, \apj, 717, 289

\bibitem[{{Steidel} {et~al.}(2001){Steidel}, {Pettini}, \&
  {Adelberger}}]{steidel01}
{Steidel}, C.~C., {Pettini}, M., \& {Adelberger}, K.~L. 2001, \apj, 546, 665

\bibitem[{{Steidel} {et~al.}(2004){Steidel}, {Shapley}, {Pettini},
  {Adelberger}, {Erb}, {Reddy}, \& {Hunt}}]{steidel04}
{Steidel}, C.~C., {Shapley}, A.~E., {Pettini}, M., {et~al.} 2004, \apj, 604,
  534

\bibitem[{{Steidel} {et~al.}(2014){Steidel}, {Rudie}, {Strom}, {Pettini},
  {Reddy}, {Shapley}, {Trainor}, {Erb}, {Turner}, {Konidaris}, {Kulas}, {Mace},
  {Matthews}, \& {McLean}}]{steidel14}
{Steidel}, C.~C., {Rudie}, G.~C., {Strom}, A.~L., {et~al.} 2014, ArXiv
  e-prints, arXiv:1405.5473

\bibitem[{{Stinson} {et~al.}(2012){Stinson}, {Brook}, {Prochaska}, {Hennawi},
  {Shen}, {Wadsley}, {Pontzen}, {Couchman}, {Quinn}, {Macci{\`o}}, \&
  {Gibson}}]{stinson12}
{Stinson}, G.~S., {Brook}, C., {Prochaska}, J.~X., {et~al.} 2012, \mnras, 425,
  1270

\bibitem[{{Stocke} {et~al.}(2006){Stocke}, {Penton}, {Danforth}, {Shull},
  {Tumlinson}, \& {McLin}}]{stocke06}
{Stocke}, J.~T., {Penton}, S.~V., {Danforth}, C.~W., {et~al.} 2006, \apj, 641,
  217

\bibitem[{{Tepper-Garc{\'{\i}}a} {et~al.}(2011){Tepper-Garc{\'{\i}}a},
  {Richter}, {Schaye}, {Booth}, {Dalla Vecchia}, {Theuns}, \&
  {Wiersma}}]{tepper-garcia11}
{Tepper-Garc{\'{\i}}a}, T., {Richter}, P., {Schaye}, J., {et~al.} 2011, \mnras,
  413, 190

\bibitem[{{Trainor} \& {Steidel}(2012)}]{trainor12}
{Trainor}, R.~F., \& {Steidel}, C.~C. 2012, \apj, 752, 39

\bibitem[{{Tripp} {et~al.}(2008){Tripp}, {Sembach}, {Bowen}, {Savage},
  {Jenkins}, {Lehner}, \& {Richter}}]{tripp08}
{Tripp}, T.~M., {Sembach}, K.~R., {Bowen}, D.~V., {et~al.} 2008, \apjs, 177, 39

\bibitem[{{Tumlinson} {et~al.}(2011){Tumlinson}, {Thom}, {Werk}, {Prochaska},
  {Tripp}, {Weinberg}, {Peeples}, {O'Meara}, {Oppenheimer}, {Meiring}, {Katz},
  {Dav{\'e}}, {Ford}, \& {Sembach}}]{tumlinson11}
{Tumlinson}, J., {Thom}, C., {Werk}, J.~K., {et~al.} 2011, Science, 334, 948

\bibitem[{{Turner} {et~al.}(2014){Turner}, {Schaye}, {Steidel}, {Rudie}, \&
  {Strom}}]{turner14}
{Turner}, M.~L., {Schaye}, J., {Steidel}, C.~C., {Rudie}, G.~C., \& {Strom},
  A.~L. 2014, \mnras, 445, 794

\bibitem[{{van de Voort} \& {Schaye}(2012)}]{vandevoort12}
{van de Voort}, F., \& {Schaye}, J. 2012, \mnras, 423, 2991

\bibitem[{{van de Voort} {et~al.}(2011){van de Voort}, {Schaye}, {Booth},
  {Haas}, \& {Dalla Vecchia}}]{vandevoort11a}
{van de Voort}, F., {Schaye}, J., {Booth}, C.~M., {Haas}, M.~R., \& {Dalla
  Vecchia}, C. 2011, \mnras, 414, 2458

\bibitem[{{Werk} {et~al.}(2013){Werk}, {Prochaska}, {Thom}, {Tumlinson},
  {Tripp}, {O'Meara}, \& {Peeples}}]{werk13}
{Werk}, J.~K., {Prochaska}, J.~X., {Thom}, C., {et~al.} 2013, \apjs, 204, 17

\bibitem[{{Werk} {et~al.}(2014){Werk}, {Prochaska}, {Tumlinson}, {Peeples},
  {Tripp}, {Fox}, {Lehner}, {Thom}, {O'Meara}, {Ford}, {Bordoloi}, {Katz},
  {Tejos}, {Oppenheimer}, {Dav{\'e}}, \& {Weinberg}}]{werk14}
{Werk}, J.~K., {Prochaska}, J.~X., {Tumlinson}, J., {et~al.} 2014, \apj, 792, 8

\end{thebibliography}


\appendix

\section{Correction of OVI contamination}
\label{sec:o6_test}

\begin{figure*}
 \includegraphics[width=\wc]{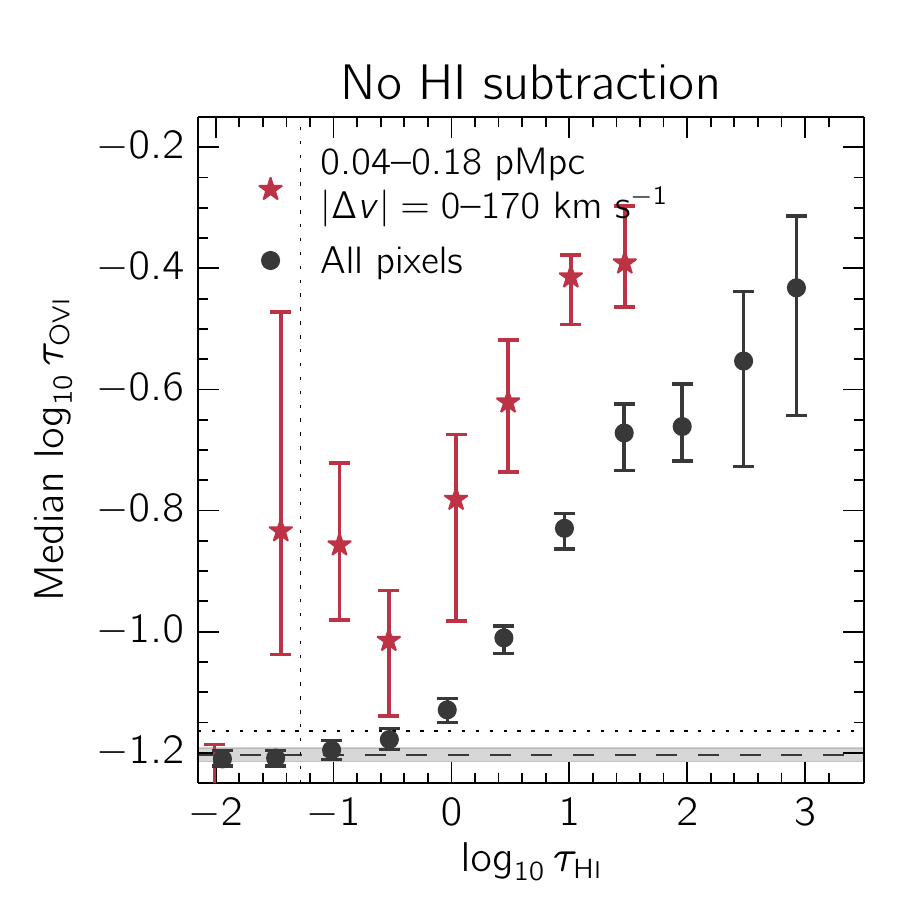} 
  \includegraphics[width=\wc]{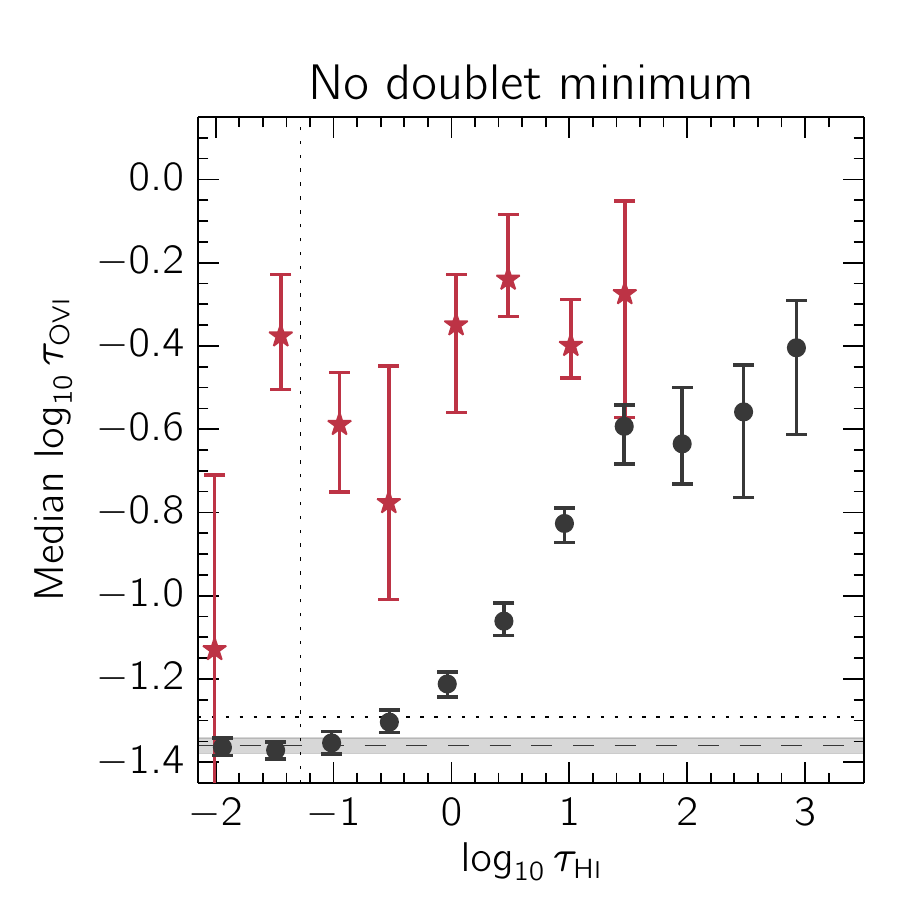} 
 \includegraphics[width=\wc]{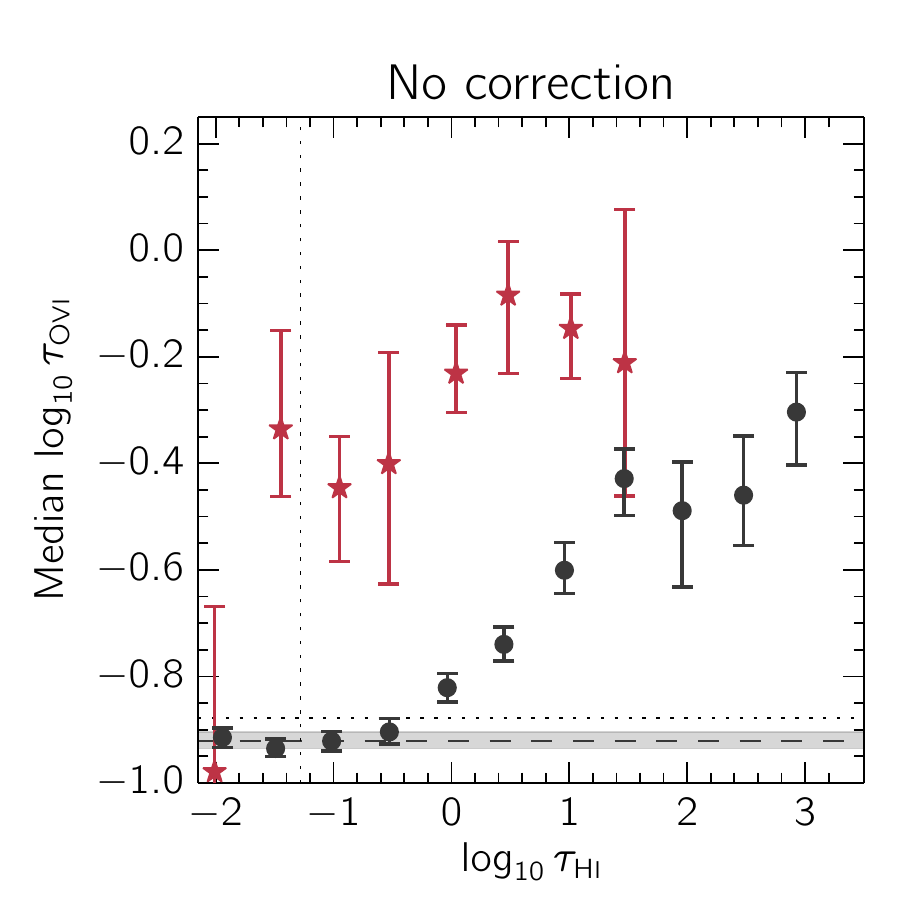} \\
 \caption{The same as the left panel Fig.~\ref{fig:o6_vs_h1_a}, but varying the method
  of \osix\ contamination correction. The modifications shown here are the result of not subtracting
  any \hone\ (left panel), not taking the doublet minimum (centre panel), and not performing
  any correction at all (right panel). This figure demonstrates that the qualitative behaviour
  of the enhancement of \osix\ optical depth at fixed \hone\ for pixels at small galactocentric
  distance is entirely independent of the \osix\ contamination correction procedure. 
}
 \label{fig:o6_test}
\end{figure*}

Here we examine how modifying the \osix\ contamination correction affects 
the resulting \osix(\hone) relation. It is possible that the small galactocentric distance
pixel sample suffers from more \hone\ contamination than the full pixel sample, given the 
larger on average \hone\ optical depth values (see the left panel of  Figure~\ref{fig:odhist}) and the proximity of 
\lyb\ to \osix\ ($\sim1800$~\kmps). We postulate that if the enhancement of \osix\ for fixed \hone\ optical depths at 
small galactocentric distances were largely due to uncorrected \hone\ \lyb\ contamination, modification 
of the \osix\ contamination would have a notable effect on this enhancement. 

To test this, in Fig.~\ref{fig:o6_test} we show the \osix(\hone) relation as in 
the left panel of Fig.~\ref{fig:o6_vs_h1_a}, but with changes to the \osix\ contamination correction.
In the fiducial case, we subtract 5 orders of the Lyman series of \hone\ beginning with \lyb, 
and then take the minimum of the \osix\ doublet optical depths.
In figure~\ref{fig:o6_test}, we first show \osix(\hone) after not subtracting any \hone\ but still 
taking the doublet minimum (left panel), not taking the doublet minimum but still subtracting \hone\ (centre panel), 
and not performing any correction at all (right panel). Although the absolute values of the apparent \osix\ optical
depths increase when fewer contamination corrections are performed, the significance of \osix\ optical depth at 
fixed \hone\ for pixels at small galactocentric distance compared to the full pixel sample remains unchanged,
indicating that our results are not sensitive to the \osix\ contamination 
correction procedure. 

Finally, we perform a simple calculation to demonstrate that the \hone\ \lyb\ contamination 
is not predicted to have a large effect. To begin, we would like to estimate the typical
\lyb\ strength at a distance $\sim1800$~\kmps\ away from galaxies. For this, we use our measurement of 
the median optical depth as a function of distance along the galaxy LOS, as was done in
the right panels of Fig.~6 from \citet{turner14}, but extending out to $\sim1800$~\kmps. 
We find that the \hone\ \lya\ optical depths asymptote to the median value of all pixels (0.051), 
and using the relative oscillator strengths of \lya\ and \lyb\ we can convert this to an expected median value
of \hone\ \lyb\ (0.0082). Adding this to the median value of all \osix\ pixels (0.023), we obtain 0.031.
This corresponds to $\sim-1$ in the left panel of Figure~\ref{fig:o6_vs_h1_a}, and is significantly 
less than the observed \osix\ optical depth enhancement.

\section{OVI(CIV) and OVI(SiIV)}
\label{sec:o6_vs_c4_and_si4}


\begin{figure*}
 \includegraphics[width=\wc]{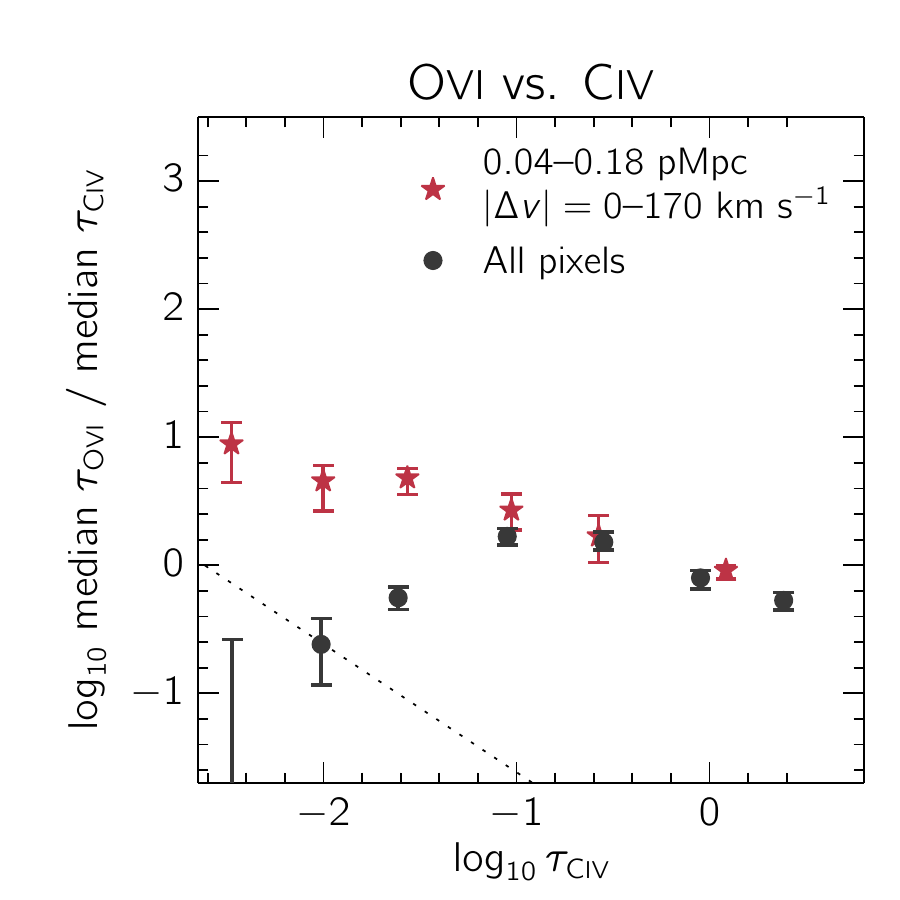} 
 \includegraphics[width=0.45\textwidth, trim=0mm 2mm 0mm 0mm]{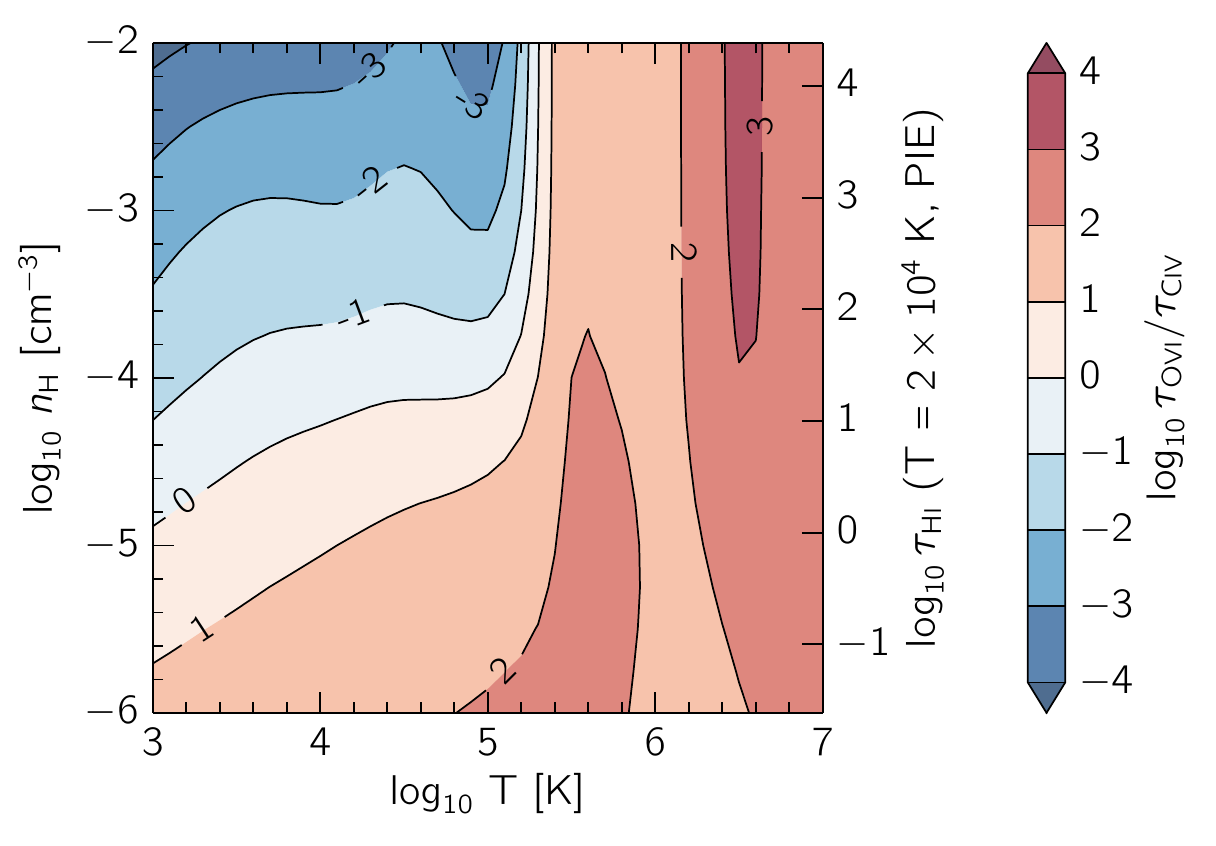}\\
 \includegraphics[width=\wc]{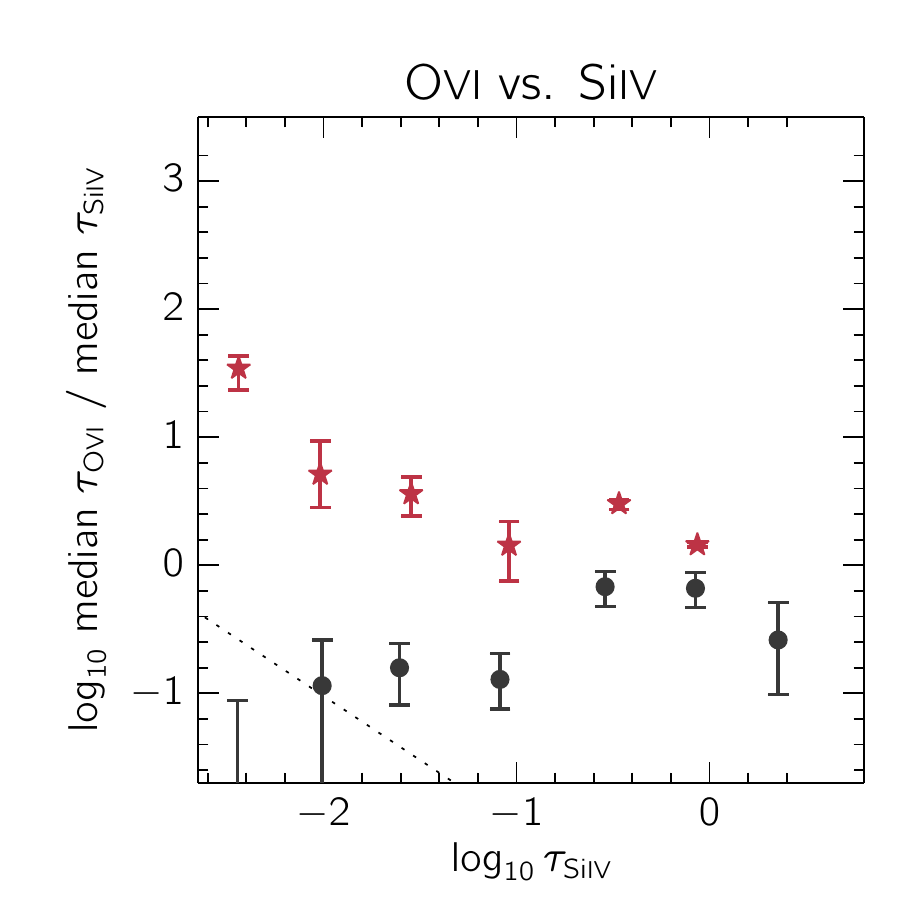} 
 \includegraphics[width=0.45\textwidth, trim=0mm 2mm 0mm 0mm]{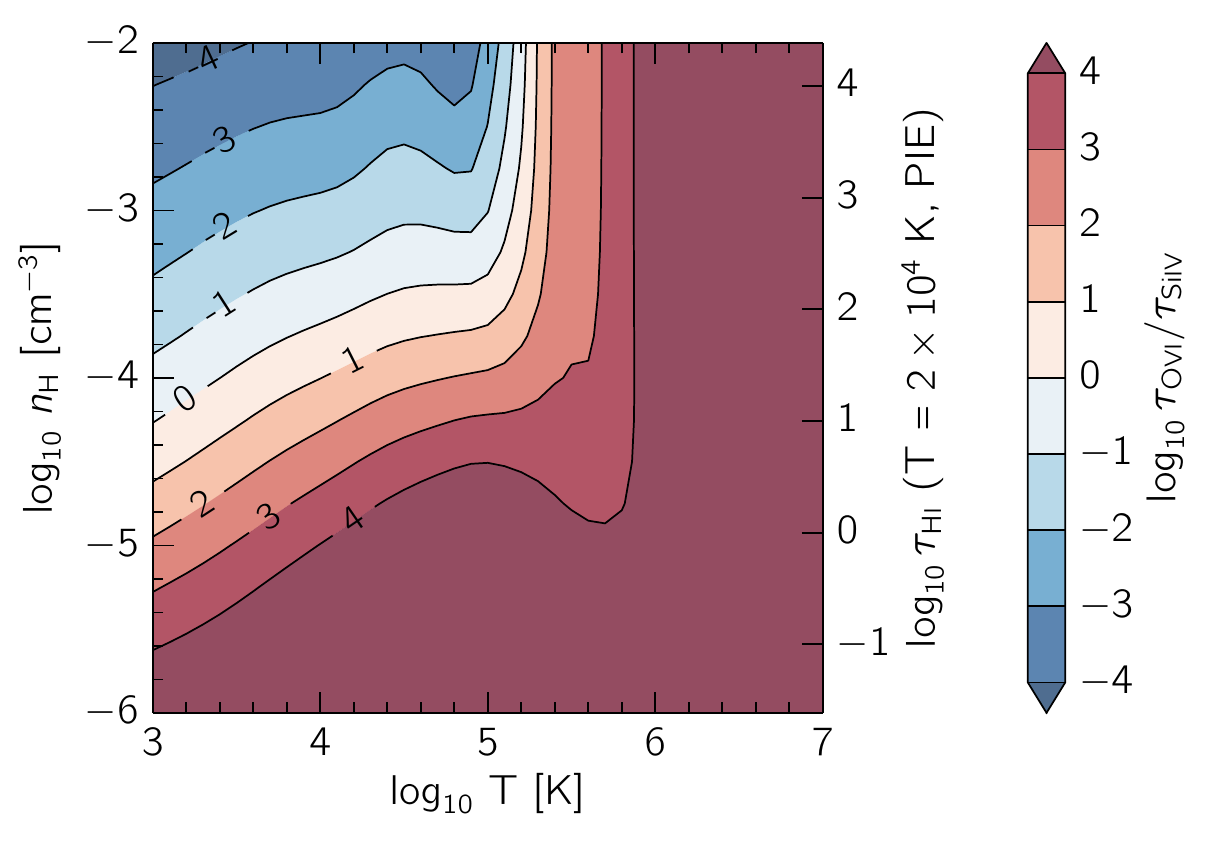}\\
 \caption{\textit{Left column:} Median optical depth ratios derived from  the \osix(\cfour) (top) 
 and \osix(\sifour) (bottom) relations in Fig.~\ref{fig:o6_vs_h1_a}.
 The diagonal lines show the median value of all \osix\ pixel optical depths 
 divided by the given value of $\tau_{\cfourm}$ and $\tau_{\sifourm}$ along the x-axis, 
   and demarcates a conservative detection level. 
  \textit{Right column:} Predicted optical depth ratios from \texttt{CLOUDY}. 
   }
 \label{fig:o6_vs_c4_and_si4_b}
\end{figure*}

In the centre and right panels of Fig.~\ref{fig:o6_vs_h1_a}, 
we investigated whether the \osix(\cfour) and \osix(\sifour) relations
depend on galactocentric distance.
First, we note that we see similar behaviour for \osix(\cfour)
 and \osix(\sifour) as for \osix(\hone) -- that is, we find a higher
 $\tau_{\osixm}$ at fixed $\tau_{\rm X}$ for pixels near galaxies
 compared to random locations, and the magnitude of this enhancement 
increases with decreasing $\tau_{\rm X}$.
 
As in \S~\ref{sec:results}, we can use \texttt{CLOUDY} to model the optical 
depth ratios as a function of temperature and density.  These
temperature-density planes, along with the observed
ratios, are shown in Fig.~\ref{fig:o6_vs_c4_and_si4_b}. 
We note that if we assume solar relative abundances, then
the optical depth ratios determined from the \texttt{CLOUDY} models are fixed 
and do not depend on metallicity (unlike for optical 
depth ratios with $\tau_{\honem}$ in the denominator).

Looking first at \osix(\cfour), for $\log_{10}\tau_{\cfourm}\gtrsim-1$
pixels near galaxies and at random locations have similar optical depth ratios ranging from
$\log_{10} \tau_{\osixm}/\tau_{\cfourm}\sim-0.4$ to $0.5$, which correspond 
to a maximum temperature of $T\sim10^{5.4}$~K. On the other hand,
at lower \cfour\ optical depths, the ratios derived from pixels near 
galaxies are as high as $\log_{10} \tau_{\osixm}/\tau_{\cfourm}\sim1$ , and are not inconsistent 
with the collisionally ionized region at $T\sim3\times10^5$~K. 

Similarly, for \osix(\sifour) the optical depth ratios derived from 
pixels with small galactocentric distances have higher values, 
and therefore higher temperature upper limits, than in the 
random regions. They reach values as high as  
$\log_{10} \tau_{\osixm}/\tau_{\sifourm}\sim1.5$, which again corresponds to a maximum temperature 
of $T\sim10^{5.4}$~K. 

Taken together, the sample of pixels at small galactocentric distance show the same
trend for \osix(\cfour) and \osix(\sifour), discrepant from that of 
random locations: $\tau_{\osixm} / \tau_{\text{x}}$ \textit{increases inversely} with
$\tau_{\cfourm}$ and $\tau_{\sifourm}$. If the conditions at near galaxies
 are most favourable to the collisional ionization of \osix\ (i.e.\ temperatures
$T=\sim\times10^{5}$~K), then we would expect to find this oxygen in regions 
with less \cfour\ and \sifour, and such a scenario is certainly not inconsistent
with the above relations.

\section{Ionization background}

\begin{figure}
\includegraphics[width=\wa]{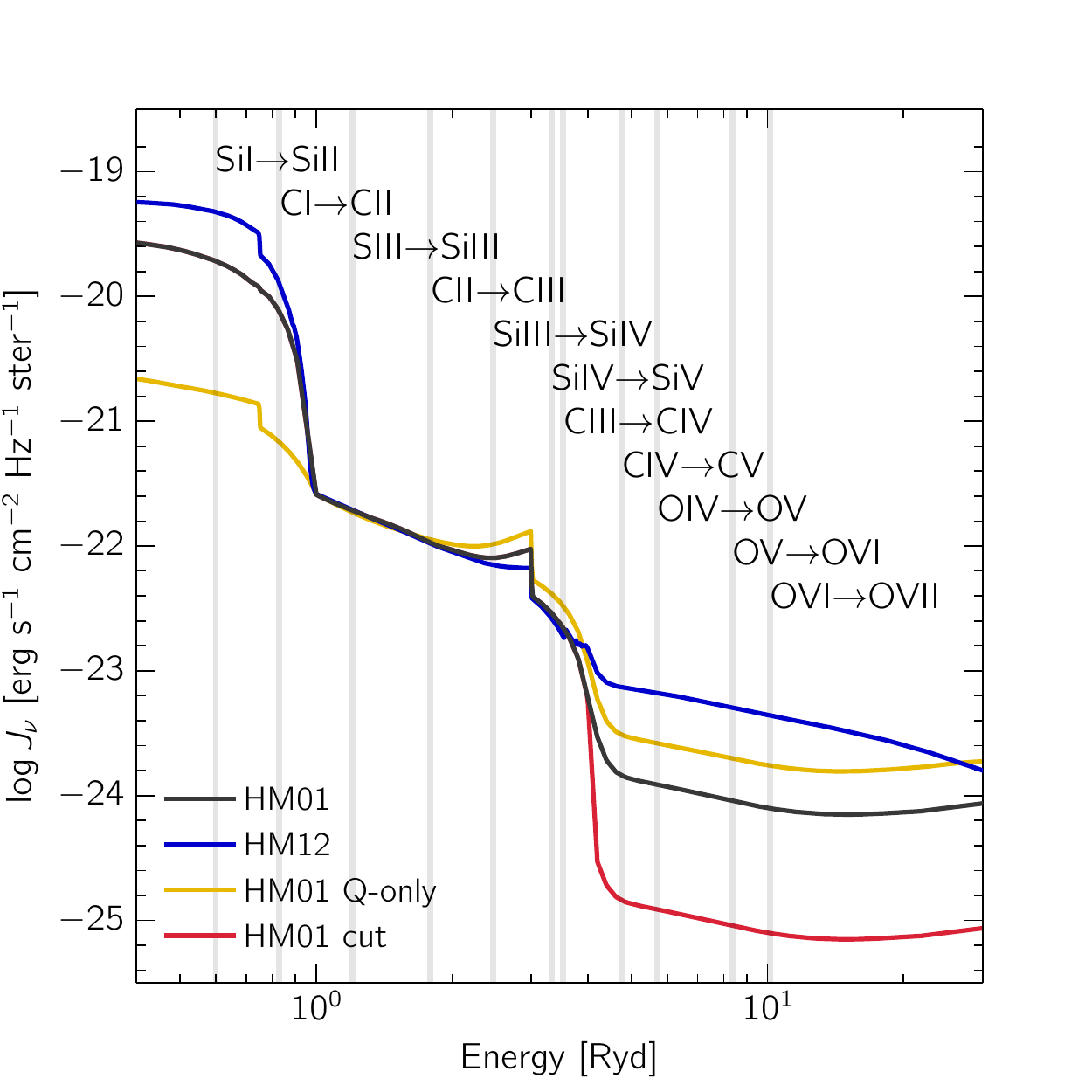} \\
\caption{Comparison of different extragalactic ionization backgrounds.}
\label{fig:bkgds}
\end{figure}

The spectral shape of the ionizing background radiation is a very large 
source of uncertainty for ionization modelling. In this section, we 
explore the impact that changing this background has on our results.
We experiment with four different models, shown in Fig.~\ref{fig:bkgds}, 
which have all been normalized to have the same \hone\ photoionization rate,
which we take from \citep{becker07} to be $\Gamma=0.74\times10^{-12}$~s$^{-1}$ 
at $z=2.34$. 
For our fiducial background, we use the \citet{haardt01} quasar+galaxy model (HM01)
and vary it in two ways. First, we use the quasar-only model, which increases the intensity
above 1~Ryd (HM01 Q-only). We also invoke a model where the
intensity above 4~Ryd is reduced by 1~dex, in order to simulate the 
absence of helium reionization (HM01 cut). Finally, we compare with the updated
background from \citet[HM12]{haardt12}.

\label{sec:bkgd}

\begin{figure*}
\includegraphics[width=\wc]{{figures/ratio_h1_o6_bins_it_velmin0_velmax169_Z_logT4.3_hm01bh_exactgamma}.pdf} 
\includegraphics[width=\wc]{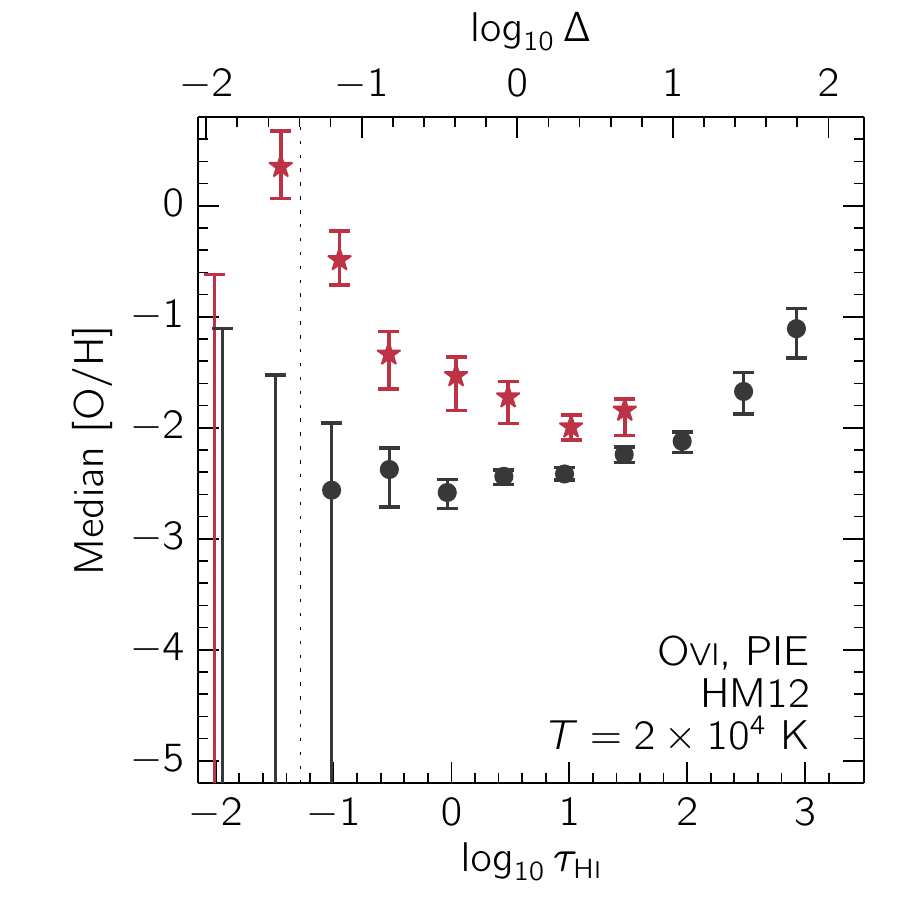} \\
\includegraphics[width=\wc]{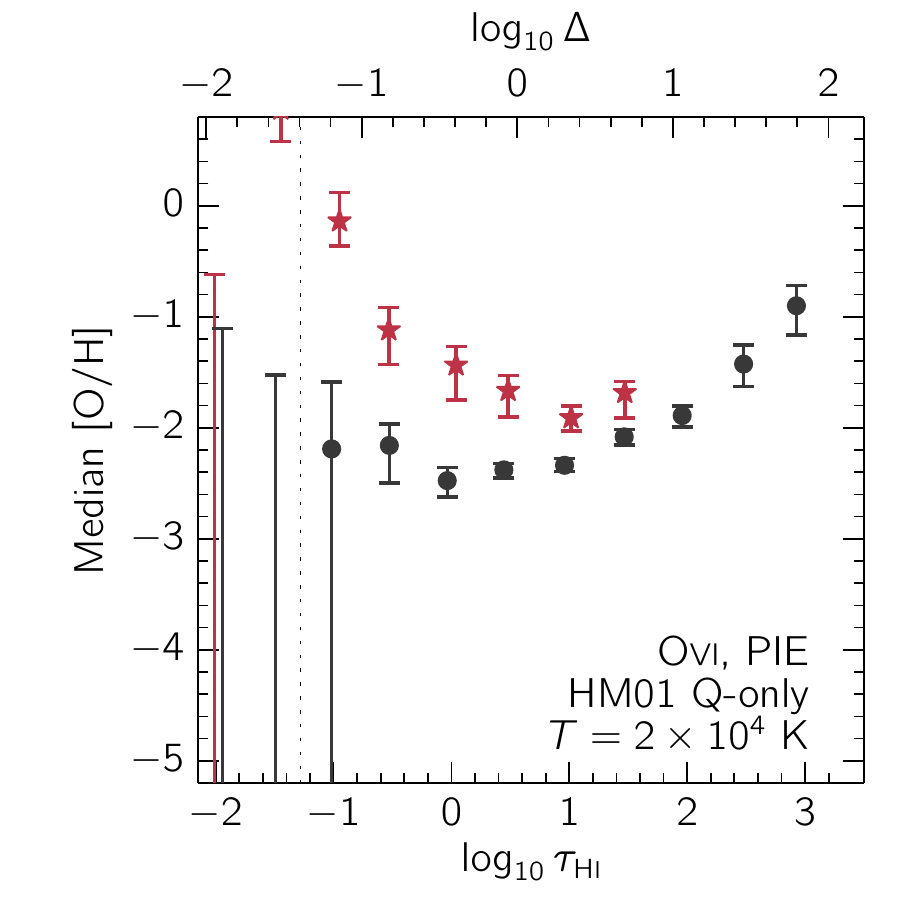} 
\includegraphics[width=\wc]{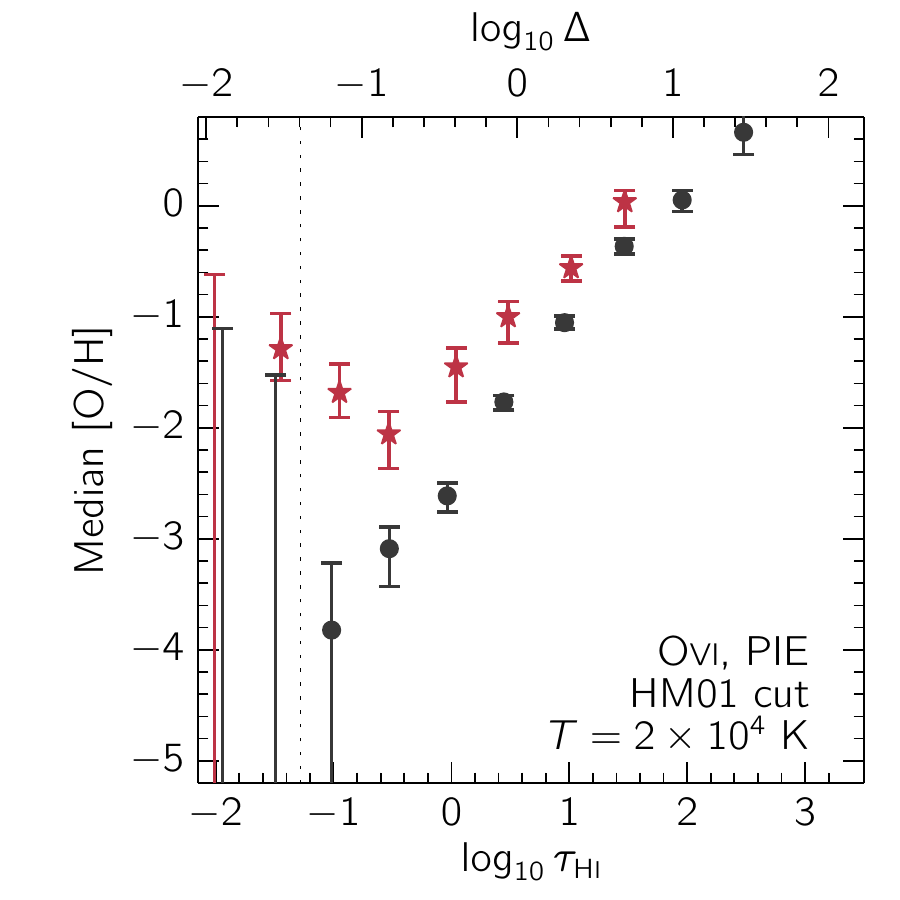} \\
\caption{[O/H] derived from the \osix(\hone) relation, for four different extragalactic backgrounds 
  (as indicated in each panel), under the assumption of PIE. 
}
\label{fig:bkgd_o6_vs_h1}
\end{figure*}

As a first test, we examine the values of [O/H] derived when we assume PIE with $T=2\times10^4$~K for the four different 
backgrounds in Fig.~\ref{fig:bkgd_o6_vs_h1}. We find that for the first three backgrounds 
(HM01, HM12, and Q-only) the results are in fairly good agreement. In general, the metallicity-density 
relation is steepest for HM01, and spans the largest range of [O/H] values.

Of the four backgrounds considered, the one salient outlier is the HM01 cut 
background (bottom-right panel). First, for the small galactocentric distance 
points, invoking this background produces reasonable (i.e.\ at least less than solar) metallicities 
for $\log_{10} \tau_{\honem} < 1$. However, the qualitative behaviour of inferred [O/H] 
as density decreases is still present.
Furthermore, for $\log_{10} \tau_{\honem} > 1$,  the inferred metallicities of the full pixel sample
become unrealistically high (up to [O/H]$\sim1$, while \citealt{steidel14} find $Z=0.4$~Z$_\odot$ 
for this galaxy sample), suggesting that this background is 
not a realistic choice. 

\begin{figure}
\begin{center}
\includegraphics[width=0.36\textwidth, trim=0mm 0mm 0mm 0mm]{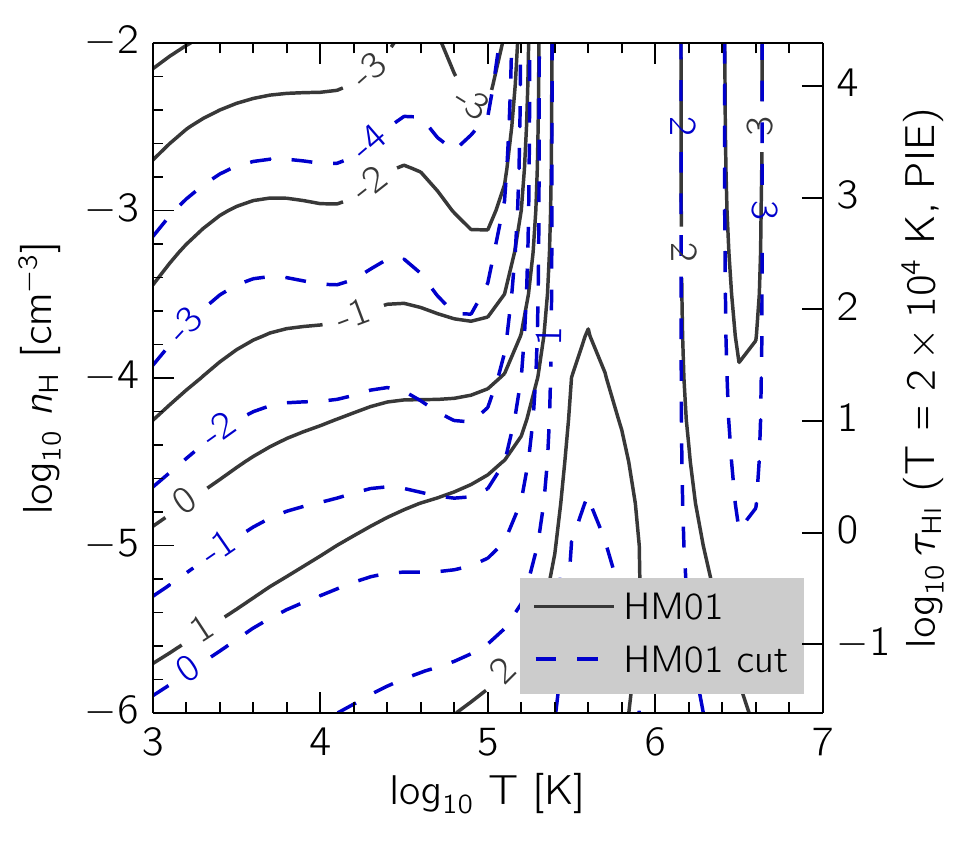}\\
\caption{The predicted values of $\log_{10} \tau_{\osixm} / \tau_{\cfourm}$
   for the HM01 (black solid contours) and HM01 cut (blue dashed contours) backgrounds.
   }
\label{fig:bkgd_o6_vs_c4}   
\end{center}
\end{figure}

Another significant change that would result from using the HM01 cut 
model is shown in Fig.~\ref{fig:bkgd_o6_vs_c4}, where we compare the 
the values of $\log_{10}\tau_{\osixm}/\tau_{\cfourm}$ 
as a function of temperature and density for both the HM01 (solid black 
contours) and HM01 cut (dashed blue contours) backgrounds. 

The observed ratios span values that range from $\sim10^0$ to $10^1$
for the small galactocentric distance sample. 
For the HM01 background, this corresponds to gas that has either 
$\log_{10} n_{\text{H}}\sim-5$ to $-4$~cm$^{-3}$ at $T\sim2\times10^4$~K, 
or $\log_{10} n_{\text{H}}\gtrsim-4.5$~cm$^{-3}$ at $T\sim10^{5.4}$~K. 
However, for HM01 cut, at the typical photoionized gas temperature
of $T\sim2\times10^4$~K, 
$\log_{10} n_{\text{H}}\lesssim-5.2$~cm$^{-3}$ for the range of observed optical depths, 
which corresponds to underdense values (i.e.\ $\Delta < 1$). 
Although the small galactocentric distance
pixel sample could be reconciled with temperatures of $T\sim10^{5.4}$~K,
it is unlikely that the full pixel sample is typically probing such hot
gas, or densities below the cosmic mean. We conclude that the HM01
cut background is in tension with ratios of  $\tau_{\osixm}/\tau_{\cfourm}$
derived from the full pixel sample.

\label{lastpage}
\end{document}